\documentclass{IEEEtran}
\usepackage{lineno}
\usepackage{amssymb}
\usepackage{amsfonts}
\usepackage{amsmath}
\usepackage{algorithm}
\usepackage{algorithmic}
\usepackage{tikz}
\usepackage{graphicx,subfigure,cite,multicol,multirow,diagbox,booktabs,array}
\usepackage{color}
\usepackage{bm}
\usepackage{subfigure}
\usepackage{stfloats}

\usetikzlibrary{shapes.geometric, arrows}
\ifCLASSINFOpdf
\else
\fi

\begin{document}
\bibliographystyle{IEEEtran}

\title{Beamforming and Phase Shift Design for HR-IRS-aided
Directional Modulation Network with a Malicious Attacker}

\author{{Feng Shu,~Hangjia He,~Rongen Dong,~Yeqing Lin,~Long Shi,\\
~Qiankun Cheng,~Jun Li and Jiangzhou~Wang,~\IEEEmembership{Fellow,~IEEE}}
\thanks{This work was supported in part by the National Natural Science Foundation of China (Nos.U22A2002, and 62071234), the Major Science and Technology plan of Hainan Province under Grant ZDKJ2021022, and the Scientific Research Fund Project of Hainan University under Grant KYQD(ZR)-21008.}
\thanks{Feng Shu is with the School of Information and Communication Engineering, Hainan University, Haikou 570228, China. and also with the School of Electronic and Optical Engineering, Nanjing University of Science and Technology, Nanjing 210094, China. (e-mail: shufeng0101@163.com. )}
\thanks{Hangjia He is with the School of Electronic and Optical Engineering, Nanjing University of Science and Technology, Nanjing, 210094, China. }
\thanks{Rongen Dong, ~Yeqing Lin and Qiankun Cheng are with the School of Information and Communication Engineering, Hainan University, Haikou 570228, China. }
\thanks{Long Shi and Jun Li are with the School of Electronic and Optical Engineering, Nanjing University of Science and Technology, Nanjing 210094, China. (e-mail: longshi@njust.edu.cn, jun.li@njust.edu.cn )}
\thanks{Jiangzhou Wang is with the School of Engineering, University of Kent, Canterbury CT2 7NT, U.K. (e-mail: j.z.wang@kent.ac.uk).}}
\maketitle

\begin{abstract}
In this paper, we propose to use hybrid relay-intelligent reflecting surface (HR-IRS) to improve the security performance of directional modulation (DM) system. In particular, the eavesdropper in this system works in full-duplex (FD) mode and he will eavesdrop on the confidential message (CM) as well as send malicious jamming. We aim to maximize the secrecy rate (SR) by jointly optimizing the receive beamforming, transmit beamforming and phase shift matrix (PSM) of HR-IRS. Since the optimization problem is un-convex and the variables are coupled to each other, we solve this problem by iteratively optimizing these variables. The receive beamforming and transmit beamforming are obtained based on  generalized Rayleigh-Ritz theorem and Dinkelbach's Transform respectively. And for PSM, two methods, called separate optimization of PSM (SO-PSM) and joint optimization of PSM (JO-PSM) are proposed. Thus, two iterative algorithms are proposed accordingly, namely maximizing SR based on SO-PSM (Max-SR-SOP) and maximizing SR based on JO-PSM (Max-SR-JOP). The former has better performance and the latter has lower complexity. The simulation results show that when HR-IRS has sufficient power budget, the proposed Max-SR-SOP and Max-SR-JOP can enable HR-IRS-aided DM network to obtain higher SR than passive IRS-aided DM network.
\end{abstract}


\section{Introduction}
In recent years, wireless communication technology develops rapidly and plays an increasingly important role in various fields. Services implemented through wireless communication emerge endlessly, and various new applications such as extended reality (XR)  also put forward more strict requirements for wireless communication technology. The implementation of XR requires that wireless systems must simultaneously provide high reliability, low latency, and high data rates \cite{6G}. Meanwhile, the application of XR involves the interaction of a large amount of private information, so the risk of privacy information disclosure caused by the broadcast nature of electromagnetic wave also needs to be paid much attention \cite{XR,pls}. Therefore, it is particularly critical for the development of XR to study how to transmit private information securely and reliably in wireless environment at high speed.

After Wyner proposed the concept of secrecy capacity, physical layer security has gained a lot of interest over the past decades \cite{wiretap,pls1}. Different from traditional upper-layer encryption technologies, physical layer security ensures the secure transmission of information at the physical layer by utilizing the physical layer characteristics of wireless communication, thus preventing information leakage and tampering. There has been extensive researches on physical layer security, among which, directional modulation (DM) is one of the key technologies. DM provides security through its directive property and is suitable for line-of-sight (LOS) channels such as millimeter wave, unmanned aerial vehicles, satellite communications, and intelligent transportation \cite{DM0}. Traditional DM generates direction-related signals by phased array or array with pattern-reconfigurable elements, so that the signal outside the desired direction is distorted, and only the desired direction can receive the undistorted signal \cite{DM1,DM11,DM2}. By combining beamforming and artificial noise (AN), DM can also be realized by baseband signal processing. In this case, it can improve the problem of sidelobe leakage and even when the receive signal power is high, illegal users will not be able to recover confidential message (CM) due to the distorted constellation diagram \cite{DM3,DM31,DM4}. In particular, the angles of legitimate and illegal users are obtained before transmission by direction of arrival (DOA) estimation \cite{doa}.

However, in the practical application of physical layer security technology, high complexity, high hardware cost and high energy consumption are still the key problems to be solved, for which the  newly emerged intelligent reflecting surface (IRS) provides a good solution \cite{irs0}. Traditional IRS consists of a large number of reconfigurable low-cost passive reflection units, each of which can independently change the amplitude or phase of the incident signal, making it possible to artificially improve the wireless propagation environment \cite{irs1}, and there has been a lot of researches on the IRS. For IRS assisted multi-input single-output (MISO) systems, the authors in \cite{irs3} and \cite{irs4} investigated energy efficiency (EE) maximization communication with IRS. And the authors in \cite{irs7} firstly proposed an alternating algorithm aiming at maximizing the secrecy rate (SR) for the scenario with single-antenna eavesdropper, which jointly optimized the transmit covariance matrix and phase shift matrix (PSM) of IRS, and then they extended the proposed algorithm to the scenario with multi-antenna eavesdropper. For IRS assisted multi-input multi-output (MIMO) system, \cite{channel} proposed a general wideband non-stationary channel model, which can be used to describe various IRS-assisted communication scenarios by adjusting parameters, while the authors in \cite{irs8} studied the basic capacity limit of the system by jointly optimizing the transmit covariance matrix and PSM of IRS, and proved that the proposed algorithm could significantly increase the channel capacity. Furthermore, \cite{irscovert} explored IRS assisted covert communications and showed the significant performance gain achieved by deploying an IRS.

Nevertheless, none of the existing works considers the presence of malicious jamming, which may occur in actual communication. In addition, since IRS also reflects jamming signals, deploying IRS can also cause harmful interference with wireless communications \cite{jamming0}. In view of the anti-jamming problem of IRS assisted system, \cite{jam0} proposed a method based on fast reinforcement learning to jointly optimize the transmit power allocation and PSM of IRS. And in \cite{jam1}, the problem of both anti-jamming and anti-eavesdropping was sovled through the joint optimization of transmit beamforming and PSM of IRS. Additionally, an anti-jamming algorithm was proposed in \cite{jam2} by jointly optimizing the coordinates and PSM of IRS. All the above studies consider the case that the receiver is equipped with a single antenna, and when the receiver is multi-antenna, the receive beamforming design is an effective anti-jamming scheme \cite{2021Low}. At present, the anti-jamming scheme combining receive beamforming and IRS PSM design is still to be studied.

Moreover, combining IRS with other technologies, such as relay \cite{irsrelay} and spatial modulation \cite{irssm}, has been proved to  further improve system performance. And the introduction of IRS into DM will also better guarantee the secure transmission in DM \cite{irsdm0}. With the assistance of IRS, the DM system can transmit two bit streams at the same time, and according to \cite{irsdm1} and \cite{irsdm2}, this scheme will greatly improve the SR performance of the DM system.

Besides, compared with relay, traditional IRS has higher EE, but due to its passive characteristic, the degree of freedom of IRS beamforming is also limited, so that IRS requires a large number of passive elements to outperform relay \cite{hrirs00}. In \cite{HR-IRS0}, a new type of IRS was proposed, namely hybrid relay-IRS (HR-IRS), which is composed of a large number of passive elements and a small number of active elements. \cite{HR-IRS0} also proved that HR-IRS had higher spectral efficiency (SE) and EE than traditional passive IRS, and in most cases, the performance of HR-IRS assistance system was better than that of relay assistance system. The authors in \cite{hrs1} also considered adding active elements to passive IRS and then used it for energy transmission, and proved that IRS with hybrid structure had better performance than traditional IRS. And in \cite{hrcovert}, the HR-IRS assisted covert communication was studied. The simulation results showed that HR-IRS is superior to traditional IRS in terms of obtaining higher covert rate.

Currently, the studies on HR-IRS is limitted, and the introduction of HR-IRS into DM is expected to improve the SR performance of the system. In addition, the problem of simultaneous anti-jamming and anti-eavesdropping in the presence of IRS and multi-antenna receivers remains to be investigated. Therefore, in this paper, we consider a HR-IRS-aided DM network with a malicious attacker, where Alice, Bob and Mallory are employed with multiple antennas. Specially, Mallory works in full-duplex (FD) mode, and will eavesdrop on CM and send malicious jamming simultaneously. Our main contributions are summarized as follows:

1) To improve the anti-jamming and anti-eavesdropping performance of the system, we introduce HR-IRS into DM, and construct a HR-IRS aided secure DM system model, where the eavesdropper Mallory works in FD mode. Under this model, we jointly optimize the receive beamforming, transmit beamforming and PSM of HR-IRS to maximize the SR of the system. In addition, considering that Mallory is a non-cooperative user and some parameters may not be obtained, we process the objective function and power constraint to make it more suitable for the actual scenario. Since the formulated optimization problem is un-convex and the variables are coupled to each other, we then solve this problem by iteratively optimizing the transmit beamforming, receive beamforming and PSM.

2) When the PSM of HR-IRS and transmit beamforming is fixed, we obtain the optimal receive beamforming according to the generalized Rayleigh-Ritz theorem. Then with fixed PSM and receive beamforming, the optimized transmit beamforming is calculated based on Dinkelbach's Transform. As for PSM, since the corresponding part of passive elements and the corresponding part of active elements are coupled to each other in the objective function and have different constraints, we find the solution of PSM by separate optimization of PSM (SO-PSM), which optimizes the passive phase shift vector (PSV) based on semi-definite relaxation (SDR) with fixed active PSV, and optimizes the active PSV by transforming the problem to a semi-definite programming (SDP) problem, with fixed passive PSV. Then, an iterative algorithm, called maximizing SR based on SO-PSM (Max-SR-SOP) is proposed accordingly.

3) To reduce the high computational complexity of the PSM design in the proposed Max-SR-SOP, a low-complexity optimization method for PSM is proposed, which is joint optimization of PSM (JO-PSM). In this method, we relax the original problem to one which maximizes the lower bound of the objective function, and then the passive PSV and active PSV are no longer coupled. In addition, in each iteration, a closed-form solution of passive PSV can be obtained, while a semi-closed-form solution of active PSV can be obtained by Lagrange multiplier method. Next, by adopting JO-PSM, the iterative algorithm called maximizing SR based on JO-PSM (Max-SR-JOP) is proposed. Compared to Max-SR-SOP, Max-SR-JOP greatly reduces the computational complexity although its performance is degraded. Moreover, simulation results show that with sufficient power budget on HR-IRS, it outperforms passive IRS in security performance, even under the same total system power budget.

The remainder of this paper is organized as follows. In Section II, we introduce the system model and formulate the optimization problem. The transmit beamforming and the receive beamforming are optimized for given PSM in Section III. And in Section IV, two PSM optimization algorithms are proposed. The overall algorithms and complexity analysis are presented in Section V, and Section VI shows the simulation results. Finally, Section VII draws our conclusion.

\emph{Notations:} In this paper, matrices, vectors, and scalars are denoted by uppercase bold, lowercase bold, and lowercase letters, respectively. Signs $(\cdot)^H$, $(\cdot)^T$, $(\cdot)^*$, $(\cdot)^+$, $tr(\cdot)$ and $\mathbb E[\cdot]$ represent the conjugate transpose, transpose, conjugate, pseudo-inverse, trace and expectation operations respectively. $\Re \{\cdot\}$ stands for the real part of a variable and $\lambda_{max}(\cdot)$ denotes the maximum eigenvalue of a matrix. $diag\{\cdot\}$ denotes the diagonal matrix corresponding to a vector.

\section{System Model and Problem Formulation}
\subsection{System Model}
\begin{figure}[ht]
\setlength{\abovecaptionskip}{-5pt}
\setlength{\belowcaptionskip}{-10pt}
\centering
\includegraphics[scale=0.5]{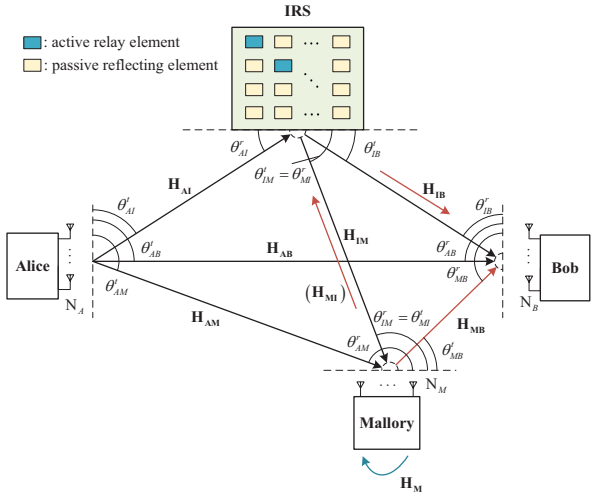}
\centering
\caption{Block diagram of HR-IRS-aided DM network with malicious attacker}
\vspace{-0.2cm}
\end{figure}

Fig.1 shows a HR-IRS-aided DM network, where the base station Alice sends CM to the legitimate user Bob with the assistance of HR-IRS, and the illegal user Mallory works in FD mode, who will not only eavesdrop on CM, but also send malicious jamming to interfere with the received signal of Bob. In this system, Alice, Bob and Mallory are equipped with $N_A$, $N_B$ and $N_M$ antennas respectively, and the HR-IRS considered in this paper is under a fixed structure, which consists of $M$ reflecting units, including $K$ active elements and $(M-K)$ passive elments, and $K \ll M$.

The transmitted baseband signal from Alice can be expressed as
\begin{align}
    {\bf s}_A = \sqrt {\beta {P_A}} {\bf{v}}{x} + \sqrt {(1 - \beta ){P_A}} {\bf T}_{A,AN}{\bf z}_{A,AN},
\end{align}
where $P_A$ denotes the transmit power of Alice, $\beta \in [0, 1]$ and $(1-\beta)$ are the power allocation parameters of CM and AN respectively. ${\bf v} \in \mathbb{C}^{N_A\times 1}$ denotes the transmit beamforming vector with the nature of ${\bf v}^H {\bf v} = 1$, while ${\bf T}_{A,AN} \in \mathbb{C}^{N_A\times N_A}$ represents the projection matrix of AN, which meets the condition $tr({\bf T}_{A,AN}{\bf T}_{A,AN}^H)=1$. $x$ is the transmitted symbol satisfying ${\mathbb E}[|x|^2]=1$, and ${\bf z}_{A,AN} \in \mathbb{C}^{N_A \times 1}$ represents the AN vector with distribution ${\bf z}_{A,AN} \sim {\mathcal {CN}}({\bf0, I}_{N_A})$.

In addition, to avoid AN interfering with the signal received by Bob,  the projection matrix of AN is designed according to the principle of null space projection, and its expression is given by
\begin{align}
    {{\bf{T}}_{A,AN}} = {{\bf{I}}_{{N_A}}} - {\bf{H}}_{CM}^H{[{{\bf{H}}_{CM}}{\bf{H}}_{CM}^H]^ + }{{\bf{H}}_{CM}},
\end {align}
where ${\bf H}_{CM} = [{\bf H}_{AI}\ {\bf H}_{AB}]^H$ represents the equivalent channel matrix of CM.

The malicious jamming signal from Mallory can be expressed as
\begin{align}
    {\bf s}_M = \sqrt {{P_M}} {\bf T}_{M,AN}{\bf z}_{M,AN},
\end{align}
where $P_M$ denotes the transmit power of Mallory, ${\bf T}_{M,AN} \in \mathbb{C}^{N_M\times N_J}$ represents the projection of jamming, which meets the condition $tr({\bf T}_{M,AN}{\bf T}_{M,AN}^H)=1$. $N_J \in [1, N_M-1]$ is the number of antennas transmitting jamming symbols, and
${\bf z}_{M,AN} \sim {\mathcal {CN}} {({\bf 0,I}_{N_J})}$ indicates the jamming symbols from Mallory.

For HR-IRS, the PSM can be written as $diag\{\theta_1, \theta_2,...,\theta_M \}$, and giving $\mathbb{Q}$ as the set of active elements in HR-IRS, $\theta_m$ is expressed as
\begin{align}
    \theta_m =
     \begin{cases}
        |a_m|e^{j \mu_m}, \quad &m \in \mathbb{Q}, \\
        e^{j \mu_m}, \quad \quad &m \notin \mathbb{Q}.
     \end{cases}
\end{align}
where $a_m$ is the amplification factor of the signal reflected by active element, and $\mu_m$ represents the phase shift of signal. For the convenience of expression, the PSM can be regarded as the sum of two matrices, i.e. $\boldsymbol \Theta = \boldsymbol \Phi + \boldsymbol \Psi$, in which $\boldsymbol \Phi = diag\{\phi_1, \phi_2,..., \phi_N \}$ and $\boldsymbol \Psi = diag\{\psi_1, \psi_2,..., \psi_N \}$ are the phase shift matrix corresponding to passive elements and active elements respectively. $\phi_n$ and $\psi_n$ are expressed as
\begin{align}
    \phi_n &=
     \begin{cases}
        0, \quad &n \in \mathbb{Q}, \\
        e^{j \mu_n}, \quad \quad &n \notin \mathbb{Q}.
     \end{cases}\\
    \psi_n &=
     \begin{cases}
        |a_n|e^{j \mu_n}, \quad &n \in \mathbb{Q}, \\
        0, \quad \quad &n \notin \mathbb{Q}.
     \end{cases}
\end{align}

Assuming that HR-IRS only reflects the signal once, the received signal at Bob can be written as
\begin{align}\label{eq0rb}
    {\bf r}_B &= {\bf v}_{BR}^H [(\underbrace{\sqrt {g_{AIB}}
        {\bf H}_{IB}^H{\boldsymbol \Theta}{\bf H}_{AI}^H
        + \sqrt {{g_{AB}}} {\bf H}_{AB}^H}_{{\bf H}_{A1}}){\bf s}_A\nonumber\\
    &+ (\underbrace{{\sqrt {g_{MIB}} {\bf H}_{IB}^H}
        {\boldsymbol \Theta}{\bf H}_{MI}^H
        + \sqrt {g_{MB}} {\bf H}_{MB}^H}_{{\bf H}_{M1}}){\bf s}_M\nonumber\\
    &+ \sqrt{g_{IB}}{\bf H}_{IB}^H {\boldsymbol \Psi}{\bf n}_{R}
        +{\bf n}_B ]\nonumber\\
    &= {{\bf v}_{BR}^H}[\sqrt {\beta {P_A}}{{\bf H}_{A1}}{\bf{v}}x
        +  \sqrt {{P_M}} {\bf H}_{M1} {\bf T}_{M,AN} {\bf z}_{M,AN}\nonumber\\
    &+ \sqrt{g_{IB}}{\bf H}_{IB}^H {\boldsymbol \Psi}{\bf n}_{R} +{\bf n}_B],
\end{align}
where ${\bf v}_{BR} \in \mathbb{C}^{N_B \times 1}$ is the receive beamforming vector of Bob, ${\bf n}_B \in \mathbb{C}^{N_B \times 1}$ denotes the complex additive white Gaussian noise (AWGN) vector with distribution ${\bf n}_B \sim {\mathcal {CN}}({\bf 0},\sigma_B^2 {\bf I}_{N_B})$. ${{\bf{n}}_R} \in \mathbb{C}^{M \times 1}$ consists of two parts, namely, the noise and the residual self-interference of the active elements, adopting the noise and self-interference model for HR-IRS in \cite{HR-IRS0}, setting $\sigma_R^2$ as the sum of noise variance and self-interference variance of active element, for each unit in ${\bf n}_R$, $n_{R,i}$ follows the distribution ${n_{R,i}}\sim{\mathcal {CN}}(0,\sigma_R^2)$ for $i \in \mathbb{Q}$, and when $i \notin \mathbb{Q}$, $n_{R,i} = 0$. $g_{AIB}$, $g_{AB}$, $g_{MIB}$, and $g_{MB}$ denote the path loss coefficients of four path: Alice to Bob through HR-IRS, Alice to Bob, Mallory to Bob through HR-IRS and Mallory to Bob.

${\bf H}_{IB}^H \in \mathbb{C}^{N_B \times M}$, ${\bf H}_{AI}^H \in \mathbb{C}^{M \times N_A}$, ${\bf H}_{AB}^H \in \mathbb{C}^{N_B \times N_A}$, ${\bf H}_{MI}^H \in \mathbb{C}^{M \times N_M}$, and ${\bf H}_{MB}^H \in \mathbb{C}^{N_B \times N_M}$ denote the channel matrices of HR-IRS to Bob, Alice to HR-IRS, Alice to Bob, Mallory to HR-IRS and Mallory to Bob respectively. And for convenience, we set ${\bf H}_{A1}$ and ${{\bf H}_{M1}}$ as the equivalent channel matrices of Alice to Bob and Mallory to Bob. Specially, in DM, there exists LOS channels, and transmitter or receiver is deployed with $N$-element linear antenna arrays. The normalized steering vector is given by
\begin{align}
    {\bf{h}}(\theta ) = \frac{1}{{\sqrt N }}
        {\left[ {{e^{j2\pi {\Psi _\theta }(1)}},...,{e^{j2\pi {\Psi _\theta }(n)}},...,{e^{j2\pi {\Psi _\theta }(N)}}} \right]^{T}},
\end{align}
where ${\Psi _\theta }(n) =  - (n - \frac{{N + 1}}{2})\frac{{d\cos \theta }}{\lambda }, n = 1...N$, with $n$ representing the index of antenna and $\theta$ representing the angle of arrival or departure of signal. $\lambda$ denotes the wavelength  and $d$ is the antenna interval. Then, the channel matrix can be given by ${\bf H}^H (\theta) = {\bf h}(\theta_r){\bf h}^H (\theta_t)$.

The received signal at Mallory can be written as
\begin{align}\label{eq0rm}
    &{{\bf r}_M} ={{\bf v}_{MR}^H}[(\underbrace{\sqrt {{g_{AIM}}}
        {{\bf H}_{IM}^H} {\boldsymbol \Theta }{\bf H}_{AI}^H
        + \sqrt {{g_{AM}}} {\bf H}_{AM}^H}_{{\bf H}_{A2}}){\bf s}_A\nonumber\\
    &+\sqrt {\rho} {\bf H}_M^H {\bf s}_M
        + \sqrt{g_{IM}}{\bf H}_{IM}^H{\boldsymbol \Psi} {\bf n}_R
        + {\bf n}_M]\nonumber\\
    &= {\bf v}_{MR}^H[\sqrt {\beta {P_A}} {\bf H}_{A2}
        {\bf{v}}x + \sqrt {(1 - \beta ){P_A}}
        {\bf H}_{A2}{\bf T}_{A,AN} {\bf z}_{A,AN}\nonumber\\
    &+ \sqrt {{\rho} {P_M}} {\bf H}_M^H {\bf T}_{M,AN}
        {\bf z}_{M,AN} + \sqrt{g_{IM}}{\bf H}_{IM}^H{\boldsymbol \Psi} {\bf n}_R
        + {\bf n}_M],
\end{align}
where ${\bf v}_{MR} \in \mathbb{C}^{N_M \times 1}$ is the receive beamforming vector of Mallory, ${\bf n}_M \in \mathbb{C}^{N_M \times 1}$ denotes the AWGN vectors, which follows the distribution ${\bf n}_M\sim {\mathcal {CN}}({\bf 0},\sigma_M^2 {\bf I}_{N_M})$. $g_{AIM}$ and $g_{AM}$ denote the path loss coefficient of the path from Alice to Mallory through HR-IRS and the direct path from Alice to Mallory respectively. ${\bf H}_{AM}^H$ denotes the channel matrix from Alice to Mallory, while ${\bf H}_{A2}$ describes the equivalent channel matrix of Alice to Mallory for convenience. Specially, due to the existence of HR-IRS and FD characteristic of Mallory, he will be subject to two types of self-interference in this model, including the self-interference between the transceiver antennas and the transmitted malicious jamming reflected by HR-IRS. Mallory will conduct self-interference cancellation for these two parts. In (\ref{eq0rm}), $\sqrt {\rho} {\bf H}_M^H {\bf s}_M$ denotes residual self-interference of Mallory, where ${\bf H}_M \in \mathbb{C}^{N_M \times N_M}$ represents the loop interference channel, and $\rho \in [0,1]$ implies the residual self-interference coefficient.

The received signal at HR-IRS can be written as
\begin{align}
{{\bf{r}}_R} &= \underbrace {\sqrt {{g_{IM}}{P_M}} {{\bf{H}}_{IM}}{{\bf{T}}_{M,AN}}{{\bf{z}}_{M,AN}}}_{{{\bf{r}}_{RM}}} + {{\bf{n}}_R}\nonumber\\
&+\underbrace {\sqrt {{g_{AI}}\beta {P_A}} {\bf{H}}_{AI}^H{\bf{v}}x}_{{{\bf{r}}_{RA}}},
\end{align}
where $g_{AI}$ and $g_{IM}$ denote the path loss coefficient of the path from Alice to HR-IRS and the path from Mallory to HR-IRS. Accordingly, when the PSM is given, the required power of HR-IRS is
\begin{align}\label{PR}
{P_R} &= \{ tr[{\bf{\Psi }}({{\bf{r}}_{RA}} + {{\bf{r}}_{RM}} + {{\bf{n}}_R}){({{\bf{r}}_{RA}} + {{\bf{r}}_{RM}} + {{\bf{n}}_R})^H}{{\bf{\Psi }}^H}]\} \nonumber\\
 &= tr[{\bf{\Psi }}({g_{AI}}\beta {P_A}{\bf{H}}_{AI}^H{\bf{v}}{{\bf{v}}^H}{{\bf{H}}_{AI}} + \sigma _R^2{{\bf{I}}_M} \nonumber\\ &+{g_{IM}}{P_M}{\bf{H}}_{MI}^H{{\bf{T}}_{M,AN}}{\bf{T}}_{M,AN}^H{{\bf{H}}_{MI}}){{\bf{\Psi }}^H}].
\end{align}

For the sake of simplicity, setting
\begin{align}
    &{\bf R}_{AB} = \beta P_A {\bf H}_{A1}{\bf v} {\bf v}^H
        {\bf H}_{A1}^H, \\
    &{\bf R}_{MJ} =  P_M {\bf H}_{M1} {\bf T}_{M,AN}
        {\bf T}_{M,AN}^H {\bf H}_{M1}^H, \\
    &{\bf R}_{BR} = \sigma_R^2g_{IB}{\bf H}_{IB}^H {\boldsymbol \Psi}
        {\boldsymbol \Psi}^H{\bf H}_{IB}\\
    &{\bf R}_{AM} = \beta P_A {\bf H}_{A2} {\bf v} {\bf v}^H
        {\bf H}_{A2}^H, \\
    &{\bf R}_{AJ} = (1-\beta)P_A{\bf H}_{A2}{\bf T}_{A,AN}
        {\bf T}_{A,AN}^H {\bf H}_{A2}^H \nonumber\\
    &\quad ~~~ = (1-\beta)P_A g_{AM} {\bf H}_{AM}^H {\bf T}_{A,AN}
            {\bf T}_{A,AN}^H {\bf H}_{AM}, \\
    &{\bf R}_{MR} = \sigma_R^2g_{IM}{\bf H}_{IM}^H {\boldsymbol \Psi}
        {\boldsymbol \Psi}^H{\bf H}_{IM}\\
    &{\bf R}_{MF} = \rho P_M {\bf H}_M^H {\bf T}_{M,AN}
        {\bf T}_{M,AN}^H {\bf H}_M,
\end{align}
the achievable rates of Bob and Mallory  are cast as
\begin{align}
    R_B = log_2(1+\frac{{\bf v}_{BR}^H {\bf R}_{AB} {\bf v}_{BR}}
        {{\bf v}_{BR}^H ({\bf R}_{MJ} + {\bf R}_{BR}) {\bf v}_{BR} + \sigma_B^2}),
\end{align}
and
\begin{align}
    R_M = log_2(1+\frac{{\bf v}_{MR}^H {\bf R}_{AM} {\bf v}_{MR}}
        {{\bf v}_{MR}^H ({\bf R}_{AJ}+{\bf R}_{MF}+{\bf R}_{MR})
        {\bf v}_{MR} + \sigma_M^2}).
\end{align}

The achievable SR is given as
\begin{align}
    R_s = {\text {max}} \{0, R_B - R_M\}.
\end{align}

\subsection{Problem Formulation}
In this paper, we aim to maximize SR by jointly optimizing the transmit beamforming $\bf v$, receive beamforming ${\bf v}_{BR}$ and PSM $\boldsymbol \Theta$. The optimization problem is formulated as
\begin{align}
    \mathrm {(P0):}~&\mathop {\max }\limits_{{\bf v},
        {\bf v}_{BR}, \boldsymbol{\Theta}}\quad \quad \quad
         \quad R_s \label{P}\\
    &\quad \ \text{s.t.} \quad \quad |{\bf v}| =1,\tag{\ref{P}{a}} \label{Pa}\\
    &\quad\quad\quad~~~~|{\bf v}_{BR}| = 1,\tag{\ref{P}{b}} \label{Pb}\\
    &\quad\quad\quad~~~~|\theta_i| = 1,
        \ \ for\  i \notin \mathbb{Q}.\tag{\ref{P}{c}} \label{Pc}\\
    &\quad\quad\quad~~~~ P_R \leq P_{R,max}\tag{\ref{P}{d}} \label{Pd}
\end{align}
where $P_{R,max}$ is the maximum power of HR-IRS.

Since in practical applications, Mallory is a non-cooperative user, it is hard to obtain some of its parameters, so further processing of problem (P0) is needed.

First, since ${\bf v}_{MR}$ is unknown to Alice and Bob in most cases, according to the generalized Rayleigh-Ritz theorem, the achievable rate of Mallory can be scaled as follows
\begin{align}
    &R_M \leq log_2( 1 + \nonumber\\
    &\quad~~~~~~~ \lambda_{max}(({\bf R}_{AJ} + {\bf R}_{MF} +{\bf R}_{MR}+ \sigma_M^2{\bf I}_{N_M})^{-1} {\bf R}_{AM}) )\nonumber\\
    & = log_2(1+ tr[({\bf R}_{AJ} + {\bf R}_{MF}+{\bf R}_{MR}
        + \sigma_M^2 {\bf I}_{N_M})^{-1} {\bf R}_{AM}]).
\end{align}

Besides, note that it is hard to obtain the residual self-interference covariance matrix of Mallory, i.e. ${\bf R}_{MF}$. We assume that Mallory completely eliminates the self-interference, as $R_M$ is upper bounded by $\overline{R}_M$, where
\begin{align}
    \overline{R}_M
     = log_2(1+ tr[({\bf R}_{AJ} +{\bf R}_{MR}
        + \sigma_M^2 {\bf I}_{N_M})^{-1} {\bf R}_{AM}]).
\end{align}

Then, considering that ${\bf T}_{M,AN}$ is also unknown, according to the compatibility of Frobenius norm, the following relationship can be derived.
\begin{align}\label{fnorm}
&\quad {\bf{v}}_{BR}^H{{\bf{R}}_{MJ}}{{\bf{v}}_{BR}} = tr[{\bf{v}}_{BR}^H{{\bf{R}}_{MJ}}{{\bf{v}}_{BR}}]\nonumber\\
&= {P_M}tr[{\bf{H}}_{M1}^H{{\bf{v}}_{BR}}{\bf{v}}_{BR}^H{{\bf{H}}_{M1}}{{\bf{T}}_{M,AN}}{\bf{T}}_{M,AN}^H]\nonumber\\
&\le {P_M}tr[{\bf{H}}_{M1}^H{{\bf{v}}_{BR}}{\bf{v}}_{BR}^H{{\bf{H}}_{M1}}]tr[{{\bf{T}}_{M,AN}}{\bf{T}}_{M,AN}^H]\nonumber\\
&= {P_M}{\bf{v}}_{BR}^H{{\bf{H}}_{M1}}{\bf{H}}_{M1}^H{{\bf{v}}_{BR}},
\end{align}
Thus, setting ${\widetilde {\bf{R}}_{MJ}} = {P_M}{{\bf{H}}_{M1}}{\bf{H}}_{M1}^H$, a lower bound of the achievable rate of Bob can be obtained as
\begin{align}
{\overline R_B} = {\log_2}(1 + \frac{{{\bf{v}}_{BR}^H{{\bf{R}}_{AB}}{{\bf{v}}_{BR}}}}{{{\bf{v}}_{BR}^H({{\widetilde {\bf{R}}}_{MJ}} + {{\bf{R}}_{BR}}){{\bf{v}}_{BR}} + \sigma_B^2}})
\end{align}

Since the power required by HR-IRS is also related to the unknown ${\bf T}_{M,AN}$, similar as (\ref{fnorm}), $P_R\leq {\overline P _R}$ can be obtained, where
\begin{align}
{\overline P _R}&= tr[{\bf{\Psi }}({g_{AI}}\beta {P_A}{\bf{H}}_{AI}^H{\bf{v}}{{\bf{v}}^H}{{\bf{H}}_{AI}} \nonumber\\
&+ {g_{IM}}{P_M}{\bf{H}}_{MI}^H{{\bf{H}}_{MI}} + \sigma _R^2{{\bf{I}}_M}){{\bf{\Psi }}^H}].
\end{align}

Consequently, the optimization problem can be transformed into
\begin{align}
    \mathrm {(P1):}~&\mathop {\max }\limits_{{\bf v},{\bf v}_{BR}, \boldsymbol{\Theta}}
         \qquad\qquad~~{ \overline R _B} - {\overline R _M} \nonumber\\
    &\quad \ \text{s.t.} \quad
        (\ref{Pa}), \ (\ref{Pb}),\  (\ref{Pc}),\  {\overline P _R} \le {P_{R,max}}.
\end{align}

As problem (P1) is non-convex and the variables are coupled with each other, it is difficult to solve it directly. Therefore, in the following, we apply the alternating optimization algorithm and optimize $\bf v$, ${\bf v}_{BR}$ and $\boldsymbol \Theta$ alternately.

\section{Beamforming Design for Given PSM}

In this section, given PSM, we first find the solution of receive beamforming with fixed transmit beamforming, and then the transmit beamforming is optimized with fixed receive beamforming.

\subsection{Optimization of ${\bf v}_{BR}$ with fixed $\bf v$ and $\boldsymbol \Theta$}

Given $\bf v$ and $\boldsymbol \Theta$, the optimization problem of ${\bf v}_{BR}$ is reduced to
\begin{align}
    \mathrm {(P2):}~&\mathop {\max }\limits_{{\bf v}_{BR} }\quad
          \frac{{\bf v}_{BR}^H {\bf R}_{AB} {\bf v}_{BR}}
          {{\bf v}_{BR}^H(\widetilde{\bf R}_{MJ}
          + {\bf R}_{BR}){\bf v}_{BR} + \sigma_B^2}, \nonumber\\
    &\ \ \text{s.t.} \quad \quad ~~~~~~~
        \quad|{\bf v}_{BR}| = 1.
\end{align}

According to the generalized Rayleigh-Ritz theorem, the optimal ${\bf v}_{BR}$ is the eigenvector corresponding to the largest eigenvalue of the matrix $(\widetilde{\bf R}_{MJ} +{\bf R}_{BR} + \sigma_B^2{\bf I}_{N_B})^{-1}{\bf R}_{AB}$.


\subsection{Optimization of $\bf v$ with fixed ${\bf v}_{BR}$ and $\boldsymbol \Theta$}

Given ${\bf v}_{BR}$ and $\boldsymbol \Theta$, the optimization problem of $\bf v$ can be transformed into
\begin{align}
    \mathrm {(P3):}~&\mathop {\max }\limits_{{\bf v}}\quad
          \frac{1 + \beta P_A \kappa^{-1}
          {\bf v}_{BR}^H {\bf H}_{A1} {\bf v}{\bf v}^H {\bf H}_{A1}^H{\bf v}_{BR}}
          {1 + tr[{\bf A}^{-1} (\beta P_A {\bf H}_{A2} {\bf v}{\bf v}^H {\bf H}_{A2}^H)]}, \nonumber\\
    &\ \ \text{s.t.} \quad \quad ~~~
        \quad |{\bf v}| =1,\ {\overline P _R} \le {P_{R,max}},
\end{align}
where
\begin{align}
    \kappa &= {\bf v}_{BR}^H (\widetilde{\bf R}_{MJ}+{\bf R}_{BR})
         {\bf v}_{BR} + \sigma_B^2,\\
    {\bf A} &= {\bf R}_{AJ} + {\bf R}_{MR} + \sigma_M^2 {\bf I}_{N_M}.
\end{align}

Besides, the upper bound of the power required by HR-IRS can be recast as
\begin{align}
{\overline P _R} &= {g_{AI}}\beta {P_A}{{\bf{v}}^H}{{\bf{H}}_{AI}}{{\bf{\Psi }}^H}{\bf{\Psi H}}_{AI}^H{\bf{v}} \nonumber\\
&+ tr[{\bf{\Psi }}({g_{IM}}{P_M}{\bf{H}}_{MI}^H{{\bf{H}}_{MI}} + \sigma _R^2{{\bf{I}}_M}){{\bf{\Psi }}^H}],
\end{align}

As a result, by setting
\begin{align}
    {\bf B} &= g_{AI} \beta P_A {\bf H}_{AI} {\boldsymbol \Psi}^H
        {\boldsymbol \Psi} {\bf H}_{AI}^H,\\
    P_{R,max}^1 &= {P_{R,max}} - tr[{\bf{\Psi }}({g_{IM}}{P_M}{\bf{H}}_{MI}^H{{\bf{H}}_{MI}} + \sigma _R^2{{\bf{I}}_M}){{\bf{\Psi }}^H}],
\end{align}
the sup-optimization problem is reformulated as
\begin{align}
    \mathrm {(P3-1):}&\mathop {\max }\limits_{{\bf v}}\
          \frac{{\bf v}^H(\beta P_A \kappa_1^{-1} {\bf H}_{A1}^H{\bf v}_{BR}{\bf v}_{BR}^H {\bf H}_{A1} + {\bf I}_{N_A}){\bf v}}
          {{\bf v}^H( \beta P_A {\bf H}_{A2}^H{\bf A}^{-1} {\bf H}_{A2} + {\bf I}_{N_A} ){\bf v}}, \nonumber\\
    &\ \ \text{s.t.} \quad \quad
        \quad |{\bf v}| = 1, \quad {\bf v}^H {\bf B}{\bf v} \leq P_{R,max}^1.
\end{align}

Since the objective function value is unsensitive to the scaling of $\bf v$, we relax the modulo constraint to ${\bf v}^H{\bf v} \leq 1$. Setting ${\bf T}_1 = \beta P_A \kappa_1^{-1} {\bf H}_{A1}^H{\bf v}_{BR}{\bf v}_{BR}^H {\bf H}_{A1} + {\bf I}_{N_A}$, ${\bf T}_2 = \beta P_A {\bf H}_{A2}^H{\bf A}^{-1} {\bf H}_{A2} + {\bf I}_{N_A}$ for brevity, the optimization problem can be recast as
\begin{align}
    \mathrm {(P3-2):}~&\mathop {\max }\limits_{{\bf v}}\quad ~~~
          \frac{{\bf v}^H {\bf T}_1{\bf v}}
          {{\bf v}^H {\bf T}_2{\bf v}},\label{V}\\
    &\ \ \text{s.t.}
        \quad~~~~ {\bf v}^H{\bf v} \leq 1,\tag{\ref{V}{a}} \label{Va}\\
    &\quad~~~~~~ \frac{{\bf v}^H {\bf B}{\bf v}}{{\bf v}^H{\bf v}} \leq P_{R,max}^1,\tag{\ref{V}{b}} \label{Vb}
\end{align}
where the power constraint (\ref{Vb}) is transformed accordingly to remove the effect of scaling $\bf v$.

Since this is a Fractional programming (FP) problem, for which, the Dinkelbach's Transform can be applied, then the optimization turns into
\begin{align}
    \mathrm {(P3-3):}~&\mathop {\max }\limits_{{\bf v}}\quad ~
          {\bf v}^H {\bf T}_1{\bf v}-
          \eta{\bf v}^H {\bf T}_2{\bf v},\label{Pv}\nonumber\\
    &\ \ \text{s.t.}
        \quad~~ (\ref{Va}),\quad(\ref{Vb})
\end{align}
where the auxiliary variable $\eta$ is updated iteratively by $\eta^{(i+1)} = {\bf v}^{(i)H} {\bf T}_1 {\bf v}^{(i)}/({\bf v}^{(i)H} {\bf T}_2 {\bf v}^{(i)})$.

However, the optimization problem (P3-3) is still non-convex, as its objective function is the difference of two convex functions and the power constraint (\ref{Vb}) is non-convex. Therefore, we then employ the successive convex approximation (SCA) method to solve the problem. By referring to the first-order Taylor expansion at fixed point ${ \tilde{\bf v}}$, the following formulas can be obtained
\begin{align}
    {\bf v}^H {\bf T}_1 {\bf v} &\geq 2 \Re\{\tilde{\bf v}^H {\bf T}_1
        {\bf v} \} -\tilde{\bf v}^H {\bf T}_1 \tilde{\bf v},\\
    {\bf v}^H {\bf v} &\geq  2 \Re\{\tilde{\bf v}^H {\bf v} \} -\tilde{\bf v}^H \tilde{\bf v},
\end{align}

Then, the optimization problem is transformed into
\begin{align}
    \mathrm {(P3-4):}~&\mathop {\max }\limits_{{\bf v}}\quad ~~
          2 \Re\{\tilde{\bf v}^H {\bf T}_1 {\bf v} \}-
          \eta{\bf v}^H {\bf T}_2{\bf v},\label{Pv}\\
    &\ \ \text{s.t.}
        \quad \quad \quad~~~ {\bf v}^H {\bf v} \leq 1,\nonumber \\
    &\quad~{\bf v}^H {\bf B}{\bf v} \leq P_{R,max}^1(2\Re\{{\bf v}^H \tilde{\bf v}\} - \tilde{\bf v}^H \tilde{\bf v}). \nonumber
\end{align}

This optimization problem is convex, and it  can be solved by CVX directly. After obtaining the solution of problem (P3-2), which is represented as $\bar{\bf v}$, the optimized transmit beamforming vector is given as ${\bf v} = \frac{\bar{\bf v}}{|\bar{\bf v}|}$. The proposed optimization method for transmit beamforming is summarized in Algorithm 1.
\begin{algorithm}
\caption{Transmit Beamformer based on Dinkelbach's Transform}
\textbf{Input:}~ ${\bf T}_1$, ${\bf T}_2$, $\bf B$, initialize ${\bf v}^{(0)}$, $\eta^{(0)}$, set $t=0$, convergence accuracy $\epsilon$.\\
\textbf{repeat:}
\begin{algorithmic}[1]
\item[1:]
    $t = t+1$.
\item[2:]
    Update $\eta^{(t)}$ as $\eta^{(t)} = \frac{{\bf{v}}^{(t-1)H} {\bf T}_1 {\bf{v}}^{(t-1)}}{\bf{v}^{(t-1)H} {\bf T}_2 \bf{v}^{(t-1)}}$.
\item[3:]
    Update ${\bf v}^{(t)}$ by setting $\widetilde {\bf{v}} = {{\bf{v}}^{(t - 1)}}$ in (P3-4) and solving problem (P3-4).
\end{algorithmic}
\textbf{until:}~${{\bf{v}}^{(t)H}}{{\bf{T}}_1}{{\bf{v}}^{(t)}} - {\eta ^{(t)}}{{\bf{v}}^{(t)H}}{{\bf{T}}_2}{{\bf{v}}^{(t)}} < \epsilon$.\\
\textbf{Output:}~${\bf{v}} = {{\bf{v}}^{(t)}}/|{{\bf{v}}^{(t)}}|$.
\end{algorithm}

\section{PSM Design for Given Beamforming}
\newcounter{TempEqCnt}
\setcounter{TempEqCnt}{\value{equation}}
\setcounter{equation}{50}
\begin{figure*}[t]
    \centering
    \begin{align}
    \overline{R}_B &= log_2(1+ \frac{({\boldsymbol \theta}^H {\bf t}_{AIB}
        + l_{AB}^*)({\bf t}_{AIB}^H{\boldsymbol \theta} + l_{AB})}
        {({\boldsymbol \theta}^H {\bf P}_{MIB}^T + {\bf t}_{MB}^T)
        ({\bf P}_{MIB}^* {\boldsymbol \theta} + {\bf t}_{MB}^*)
        + {\boldsymbol \psi}^H {\bf P}_{IB}^T {\bf P}_{IB}^* {\boldsymbol \psi} + \sigma_B^2}) \label{RB}\\
    \overline{R}_M &= log_2(1 + tr[\frac{({\bf P}_{AIM}
        {\boldsymbol \theta} + {\bf t}_{AM})({\boldsymbol \theta}^H
        {\bf P}_{AIM}^H + {\bf t}_{AM}^H)}
        {{\boldsymbol \psi}^H {\bf P}_{IM}{\boldsymbol \psi} {\bf h}_{IM,r}{\bf h}_{IM,r}^H + {\bf R}_{NJ} }])\nonumber \\
    &= log_2(1+ ({\boldsymbol \theta}^H
        {\bf P}_{AIM}^H + {\bf t}_{AM}^H) ({{\boldsymbol \psi}^H {\bf P}_{IM}{\boldsymbol \psi} {\bf h}_{IM,r}{\bf h}_{IM,r}^H + {\bf R}_{NJ}})^{-1}({\bf P}_{AIM}
        {\boldsymbol \theta} + {\bf t}_{AM})). \label{RM}
    \end{align}
    \hrulefill
\end{figure*}
\setcounter{equation}{\value{TempEqCnt}}
In this section, we first reformulate the optimization problem of PSM with fixed beamforming for the sake of simplicity, and then two optimization methods are proposed for PSM.

\subsection{Reconstruction of the subproblem}
To facilitate the subsequent calculation, we extract the phase shifts of IRS before solving the problem, and then rewrite the optimization problem of PSM. First, defining $\boldsymbol \theta$, $\boldsymbol \phi$ and $\boldsymbol \psi \in \mathbb{C}^{M\times 1}$ as the vectors formed by the diagonal elements of $\boldsymbol \Theta$, $\boldsymbol \Phi$ and $\boldsymbol \Psi$ respectively, one obtains
\begin{align}
    &\quad ~~~~~~~~~~~~~~~\sqrt{P_M}{\bf H}_{M1}^H {\bf v}_{BR}   \nonumber\\
    &=\sqrt{g_{MIB}P_M} {\bf H}_{MI} {\boldsymbol \Theta}^H
        {\bf H}_{IB} {\bf v}_{BR}
        + \sqrt{g_{MB}P_M} {\bf H}_{MB} {\bf v}_{BR} \nonumber\\
    &\overset{(a)}{=} \sqrt{g_{MIB}P_M} {\bf H}_{MI} diag\{ {\bf H}_{IB} {\bf v}_{BR} \}{\boldsymbol \theta}^* \nonumber \\
    &+ \sqrt{g_{MB}P_M} {\bf H}_{MB} {\bf v}_{BR},
\end{align}
where (a) holds due to $diag\{{\bf a}\}{\bf b} = diag\{{\bf b}\}{\bf a}$. Therefore, by setting
\begin{align}
    {\bf P}_{MIB} &= \sqrt{g_{MIB}P_M} {\bf H}_{MI} diag\{ {\bf H}_{IB} {\bf v}_{BR} \}, \nonumber \\
    {\bf t}_{MB} &= \sqrt{g_{MB}P_M} {\bf H}_{MB} {\bf v}_{BR},
\end{align}
it can be obtained that
\begin{align}
    &{\bf v}_{BR}^H {\widetilde{\bf R}_{MJ}{\bf v}_{BR}}
    =({\bf P}_{MIB} {\boldsymbol \theta}^* + {\bf t}_{MB})^H
        ({\bf P}_{MIB} {\boldsymbol \theta}^* + {\bf t}_{MB})\nonumber\\
    &~~~~~~= ({\boldsymbol \theta}^H {\bf P}_{MIB}^T + {\bf t}_{MB}^T)
        ({\bf P}_{MIB}^* {\boldsymbol \theta} + {\bf t}_{MB}^*).
\end{align}

Similarly, by setting
\begin{align}
    {\bf P}_{IB} = \sigma_{R} \sqrt{g_{IB}} diag\{ {\bf H}_{IB} {\bf v}_{BR}\},
\end{align}
it can be obtained that
\begin{align}
    {\bf v}_{BR}^H {\bf R}_{BR}{\bf v}_{BR} =
    {\boldsymbol \psi}^H {\bf P}_{IB}^T {\bf P}_{IB}^* {\boldsymbol \psi}.
\end{align}

Furthermore, we define ${\bf t}_{AIB}$, $l_{AB}$, ${\bf P}_{AIM}$ and ${\bf t}_{AM}$ as follows
\begin{align}
    {\bf t}_{AIB} &= \sqrt{\beta g_{AIB} P_A }
        diag\{{\bf{H}_{AI}^H {\bf v}}\})^H {\bf H}_{IB}{\bf v}_{BR},\nonumber \\
    l_{AB} &= \sqrt{\beta g_{AB} P_A}{\bf v}_{BR}^H
        {\bf H}_{AB}^H {\bf v},\nonumber\\
    {\bf P}_{AIM} &= \sqrt{\beta g_{AIM} P_A} {\bf H}_{IM}^H
        diag\{{\bf H}_{AI}^H{\bf v} \}, \nonumber \\
    {\bf t}_{AM} &= \sqrt{\beta g_{AM} P_A} {\bf H}_{AM}^H {\bf v},
\end{align}
then the following equations can be derived
\begin{align}
    {\bf v}_{BR}^H {\bf R}_{AB} {\bf v}_{BR}
    = ({\bf t}_{AIB}^H {\boldsymbol \theta} + l_{AB})
        ({\boldsymbol \theta}^H {\bf t}_{AIB} + l_{AB}^*), \nonumber\\
    {\bf R}_{AM} = ({\bf P}_{AIM} {\boldsymbol \theta} + {\bf t}_{AM})
        ({\boldsymbol \theta}^H {\bf P}_{AIM}^H + {\bf t}_{AM}^H).
\end{align}

Besides, since ${\bf H}_{IM}^H = {\bf h}_{IM,r} {\bf h}_{IM,t}^H$, by setting
\begin{align}
    {\bf P}_{IM} = \sigma_R^2 g_{IM} diag\{{\bf h}_{IM,t}\}
        diag\{{\bf h}_{IM,t}^*\},
\end{align}
it can be obtained that
\begin{align}\label{rmr}
    {\bf R}_{MR} = {\boldsymbol \psi}^H {\bf P}_{IM}
        {\boldsymbol \psi} {\bf h}_{IM,r}{\bf h}_{IM,r}^H.
\end{align}
\setcounter{equation}{52}

Thus, setting ${\bf R}_{NJ} = {\bf R}_{AJ} + \sigma_M^2 {\bf I}_{N_M}$, $\overline{R}_B$ and $\overline{R}_M$ can be rewritten as (\ref{RB}) and (\ref{RM}). The optimization problem of $\boldsymbol \theta$ is reformulated as
\begin{align}
    \mathrm {(P4):}~&\mathop {\max }\limits_{{\boldsymbol \theta}}\qquad ~~~~
          \overline{R}_B - {\overline{R}_M}, \\
    &\ \ \text{s.t.}
        \quad~~~ (\ref{Pc}),\quad{\overline P _R} \le {P_{R,max}}. \nonumber
\end{align}

\subsection{Proposed SO-PSM method}
As passive PSV $\boldsymbol \phi$ and active PSV $\boldsymbol \psi$ are coupled to each other in the objective function of problem (P4), in the following, we propose the SO-PSM method, which optimize $\boldsymbol \phi$ and $\boldsymbol \psi$ alternately.

First, for ease of presentation, let $\mathcal{I}_M^{\mathbb{Q}}$ denote the following matrix
\begin{align}
    {\mathcal{I}_M^{\mathbb{Q}}}(i,j) =
    \begin{cases}
        1, \quad i = j, and ~i \in \mathbb{Q}, \\
        0, \quad others,
    \end{cases}
\end{align}
and define $\mathcal{I}_M^{\bar{\mathbb{Q}}} = {\bf I}_M - \mathcal{I}_M^{\mathbb{Q}}$, therefore we have ${\boldsymbol \phi} = \mathcal{I}_M^{\bar{\mathbb{Q}}} {\boldsymbol \theta} =  \mathcal{I}_M^{\bar{\mathbb{Q}}} {\boldsymbol \phi}$ and ${\boldsymbol \psi} = \mathcal{I}_M^{\mathbb{Q}} {\boldsymbol \theta} = \mathcal{I}_M^{\mathbb{Q}} {\boldsymbol \psi}$. Next, with the remaining variables fixed, we solve the problems about $\boldsymbol \phi$ and $\boldsymbol \psi$ respectively.

1) Optimizing $\boldsymbol \phi$ with fixed $\boldsymbol \psi$:

When $\boldsymbol \psi$ is fixed, let us define
\begin{align}
    \tilde{l}_{AB} &= {\bf t}_{AIB}^H {\boldsymbol \psi} + l_{AB},\nonumber\\
    \tilde{\bf t}_{MB} &= {\bf P}_{MIB} {\boldsymbol \psi}^* + {\bf t}_{MB}, \nonumber\\
    \tilde{\sigma}_B^2 &= {\boldsymbol \psi}^H {\bf P}_{IB}^T {\bf P}_{IB}^* {\boldsymbol \psi} + \sigma_B^2, \\
    \widetilde{\bf R}_{NJ} &= {{\boldsymbol \psi}^H {\bf P}_{IM}{\boldsymbol \psi} {\bf h}_{IM,r}{\bf h}_{IM,r}^H + {\bf R}_{NJ}}, \nonumber\\
    \tilde{\bf t}_{AM} &= {\bf P}_{AIM} {\boldsymbol \psi} + {\bf t}_{AM},\nonumber
\end{align}
and
\begin{align}
    &\tilde{\bf P}_{MIB} =
        {\bf P}_{MIB} \mathcal{I}_M^{\bar{\mathbb{Q}}}, \
    \widetilde{\bf P}_{AIM} =
        {\bf P}_{AIM} \mathcal{I}_M^{\bar{\mathbb{Q}}},\nonumber\\
    & \qquad \qquad~\tilde{\bf t}_{AIB} = \mathcal{I}_M^{\bar{\mathbb{Q}}}{\bf t}_{AIB},
\end{align}
then the optimization problem of $\boldsymbol \phi$ can be expressed as
\begin{align}\label{phi}
    &\mathrm {(P5):}
    \mathop {\max }\limits_{{\boldsymbol \phi}} \
          \frac{1+ |\tilde{\bf t}_{AIB}^H{\boldsymbol \phi} + \tilde{l}_{AB}|^2
          / (|\widetilde{P}_{MIB}^* {\boldsymbol \phi} + \tilde{\bf t}_{MB}^*|^2 + \tilde{\sigma}_B^2)}
          {1+ ({\boldsymbol \phi}^H \widetilde{\bf P}_{AIM}^H + \tilde{\bf t}_{AM}^H) \widetilde{\bf R}_{NJ}^{-1}
          (\widetilde{\bf P}_{AIM} {\boldsymbol \phi} + \tilde{\bf t}_{AM})}, \nonumber \\
    &\ \ \text{s.t.}\quad~~~~~~~~~~~
        |{\boldsymbol \phi}_i| = 1,\quad i = 1,...,M,
\end{align}

Since the objective function of (P5) contains the quadratic terms of $\boldsymbol \phi$, and (P5) only has a unit-modular constraint, SDR algorithm can be used to solve this problem.

Letting $f_1(\boldsymbol \phi)$ denote the objective function of (\ref{phi}), the logarithm of it can be obtained as
\begin{align}
    &\quad~~ ln( f_1({\boldsymbol \phi})) \nonumber \\
    &= ln(|\widetilde{\bf P}_{MIB}^* {\boldsymbol \phi}
        + \tilde{\bf t}_{MB}^*|^2 + \tilde{\sigma}_B^2 +
        |\tilde{\bf t}_{AIB}^H{\boldsymbol \phi} + \tilde{l}_{AB}|^2 )
        \nonumber \\
    &- ln(|\widetilde{\bf P}_{MIB}^* {\boldsymbol \phi}
        + \tilde{\bf t}_{MB}^*|^2 + \tilde{\sigma}_B^2)\nonumber\\
    &-ln(1+ ({\boldsymbol \phi}^H \widetilde{\bf P}_{AIM}^H +
        \tilde{\bf t}_{AM}^H) \widetilde{\bf R}_{NJ}^{-1}
          (\widetilde{\bf P}_{AIM} {\boldsymbol \phi} + \tilde{\bf t}_{AM})).
\end{align}

Then, by setting
\begin{align}\label{W}
    {\bf W} = \left[
        {\begin{array}{*{20}{c}}
        {\boldsymbol \phi}\\
        {1}
        \end{array}} \right]
        [{\boldsymbol \phi}^H \quad 1],
\end{align}
and
\begin{align}
    {\bf L}_{M} &= \left[
        {\begin{array}{cc}
        {\widetilde{\bf P}_{MIB}^T \widetilde{\bf P}_{MIB}^*} &
            {\widetilde{\bf P}_{MIB}^T \tilde{\bf t}_{MB}^*}\\
        {\tilde{\bf t}_{MB}^T \widetilde{\bf P}_{MIB}^*} &
            {\tilde{\bf t}_{MB}^T \tilde{\bf t}_{MB}^* + \tilde{\sigma}_B^2}
        \end{array}} \right],\\
    {\bf L}_{A} &= \left[
        {\begin{array}{cc}
        {\tilde{\bf t}_{AIB}\tilde{\bf t}_{AIB}^H} &
            {\tilde{\bf t}_{AIB} \tilde{l}_{AB}}\\
        {\tilde{l}_{AB}^* \tilde{\bf t}_{AIB}^H} &
            {\tilde{l}_{AB}^* \tilde{l}_{AB}}
        \end{array}} \right],\\
    {\bf L}_{E} &= \left[
        {\begin{array}{cc}
        {\widetilde{\bf P}_{AIM}^H \widetilde{\bf R}_{NJ}^{-1} \widetilde{\bf P}_{AIM}} &
            {\widetilde{\bf P}_{AIM}^H \widetilde{\bf R}_{NJ}^{-1} \tilde{\bf t}_{AM}}\\
        {\tilde{\bf t}_{AM}^H \widetilde{\bf R}_{NJ}^{-1} \widetilde{\bf P}_{AIM}} &
            {1+ \tilde{\bf t}_{AM}^H \widetilde{\bf R}_{NJ}^{-1}\tilde{\bf t}_{AM}}
        \end{array}} \right],
\end{align}
$ln(f_1({\boldsymbol \phi}))$ can be equivalently transformed to
\begin{align}
    ln(f_1({\boldsymbol \phi})) &= g_1({\bf W})
        \nonumber\\
    &= ln(tr[({\bf L}_{M} + {\bf L}_{A}) {\bf W}])
        - ln(tr[ {\bf L}_{M} {\bf W}])\nonumber \\
    &- ln(tr[{\bf L}_{E} {\bf W}]).
\end{align}

Hence, the optimization problem can be relaxed to
\begin{align}
    \mathrm {(P5-1):}~&\mathop {\max }\limits_{\bf W} \quad \quad ~~~~~
          g_1({\bf W}) \nonumber \\
    &\ \ \text{s.t.}\quad
        rank({\bf W}) = 1, \quad {\bf W}\succeq {\bf 0}, \nonumber \\
        & \quad \quad \quad{\bf W}(i,i) = 1,\quad i = 1,...,M,
\end{align}
The objective function is un-concave as it is the difference of concave functions, therefore, majorization-minimization (MM) algorithm is applied next. According to \cite{sdr}, at any fixed point $\widetilde{\bf W}$, the following inequality holds
\begin{align}
    ln(tr[{\bf L W}]) \leq ln(tr[{\bf L} \widetilde{\bf W}])
        + tr[\frac{\bf L}{tr[{\bf L}\widetilde{\bf W}]} (\bf W - \widetilde{\bf W})].
\end{align}
As a result, a lower bound of $ g_1({\bf W})$ can be found as follows
\begin{align}
    g_1({\bf W}) &\geq ln(tr[({\bf L}_{M} + {\bf L}_{A}){\bf W}])
        - ln(tr[{\bf L}_{M} \widetilde{\bf W}]) \nonumber \\
        &- tr[\frac{{\bf L}_{M}}{tr[{\bf L}_{M} \widetilde{\bf W}]} ({\bf W} - \widetilde{\bf W})]
        - ln(tr[{\bf L}_{E} \widetilde{\bf W}]) \nonumber \\
        &- tr[\frac{{\bf L}_{E}}{tr[{\bf L}_{E}\widetilde{\bf W}]} ({\bf W} - \widetilde{\bf W})]
\end{align}

In addition, since the rank-one constraint is non-convex, the constraint can be removed first to obtain a relaxed optimization problem, as shown below
\begin{align}
     &\mathrm {(P5-2):}\nonumber\\
     &\mathop {\max }\limits_{\bf W} \
          ln(tr[({\bf L}_{M} + {\bf L}_{A}){\bf W}])
        - \frac{tr[{\bf L}_{M}{\bf W}]}{tr[{\bf L}_{M}\widetilde{\bf W}]}
        - \frac{tr[{\bf L}_{E}{\bf W}]}{tr[{\bf L}_{E}\widetilde{\bf W}]} \nonumber\\
    &\ \ \text{s.t.}
        \quad \quad \quad {\bf W}\succeq {\bf 0},
        \quad{\bf W}(i,i) = 1,\ i = 1,...,M,
\end{align}
This is a convex optimization problem, which can be solved by CVX. And then the Gaussian randomization method can be applied to find a rank-one solution. The specific steps of optimizing $\boldsymbol \phi$ with other variables fixed are given in Algorithm 2.

\begin{algorithm}
\caption{Optimization Algorithm to (P5)}
\textbf{Input:}~ ${\bf L}_M$, ${\bf L}_A$, ${\bf L}_E$, initialize ${\boldsymbol \phi}^{(0)}$, compute ${\bf W}^{(0)}$, $g_1^{(0)}$, set $t=0$, convergence accuracy $\epsilon$.\\
\textbf{repeat:}
\begin{algorithmic}[1]
\item[1:]
    $t = t+1$.
\item[2:]
    Update ${\boldsymbol \phi}^{(t)}$ by setting $\widetilde {\bf{W}} = {{\bf{W}}^{(t - 1)}}$ in (P5-2) and using Gaussian randomization method after solving problem (P5-2).
\item[3:]
    Update ${\bf W}^{(t)}$ by (\ref{W}).
\item[4:]
    Compute $g_1^{(t)}$.
\end{algorithmic}
\textbf{until:}~$g_1^{(t)} - g_1^{(t - 1)} < \epsilon$.\\
\textbf{Output:}~${\boldsymbol \phi}^{(t)}$.
\end{algorithm}

2) Optimizing $\boldsymbol \psi$ with fixed $\boldsymbol \phi$
\setcounter{TempEqCnt}{\value{equation}}
\setcounter{equation}{68}
\begin{figure*}
\begin{align}\label{f}
    f_2({\boldsymbol \psi}) =
        \frac{1+ |\bar{\bf t}_{AIB}^H{\boldsymbol \psi} + \bar{l}_{AB}|^2
        / (|\overline{\bf P}_{MIB}^* {\boldsymbol \psi} + \bar{\bf t}_{MB}^*|^2 + |\overline{\bf P}_{IB}^* {\boldsymbol \psi}|^2 +  {\sigma}_B^2)}
        {1+ ({\boldsymbol \psi}^H
        \overline{\bf P}_{AIM}^H + \bar{\bf t}_{AM}^H) ({{\boldsymbol \psi}^H \overline{\bf P}_{IM}{\boldsymbol \psi} {\bf h}_{IM,r}{\bf h}_{IM,r}^H + {\bf R}_{NJ}})^{-1}(\overline{\bf P}_{AIM}{\boldsymbol \psi} + \bar{\bf t}_{AM})}
\end{align}
\hrulefill
\end{figure*}
\setcounter{equation}{\value{TempEqCnt}}

Now, we turn to solve the problem of optimizing $\boldsymbol \psi$ with fixed $\boldsymbol \phi$, and to simplify the expression of the target problem, we define the following new variables
\begin{align}
    \bar{l}_{AB} &= {\bf t}_{AIB}^H {\boldsymbol \phi} + l_{AB},\qquad
    \bar{\bf t}_{MB} = {\bf P}_{MIB} {\boldsymbol \phi}^* + {\bf t}_{MB}, \nonumber\\
    \bar{\bf t}_{AM} &= {\bf P}_{AIM} {\boldsymbol \phi} + {\bf t}_{AM},~~~
    \bar{\bf t}_{AIB} = \mathcal{I}_M^{\mathbb{Q}}{\bf t}_{AIB}, \nonumber\\
    \overline{\bf P}_{MIB} &=
        {\bf P}_{MIB} \mathcal{I}_M^{\mathbb{Q}}, \qquad\quad
    \overline{\bf P}_{AIM} =
        {\bf P}_{AIM} \mathcal{I}_M^{\mathbb{Q}}, \nonumber\\
    \overline{\bf P}_{IB} &= {\bf P}_{IB} \mathcal{I}_M^{\mathbb{Q}},\qquad \qquad~
    \overline{\bf P}_{IM} = \mathcal{I}_M^{\mathbb{Q}} {\bf P}_{IM} \mathcal{I}_M^{\mathbb{Q}}.
\end{align}
Then the objective function of $\boldsymbol \psi$ can be re-cast as (\ref{f}).
\setcounter{equation}{69}

As for the power constraint, by setting ${{\bf{t}}_{AI}} = \sqrt {\beta {g_{AI}}{P_A}} {\bf{H}}_{AI}^H{\bf{v}}$ and ${{\bf{t}}_{MI}} = \sqrt {{g_{IM}}{P_M}{N_M}} {{\bf{h}}_{MI,r}}$, the upper bound of the transmit power of HR-IRS can be rewritten as
\begin{align}
    {\overline P _R} &= tr[{\boldsymbol \Psi}({\bf t}_{AI} {\bf t}_{AI}^H + {\bf t}_{MI} {\bf t}_{MI}^H  + \sigma_R^2 {\bf I}_M){\boldsymbol \Psi}^H]\nonumber\\
    &= {\boldsymbol \psi}^H \overline{\bf D}{\boldsymbol \psi},
\end{align}
where $\overline{\bf D} = \mathcal{I}_M^{\mathbb{Q}} {\bf D} \mathcal{I}_M^{\mathbb{Q}}$, and ${\bf{D}} = diag\{ {\bf{t}}_{AI}^*\} diag\{ {{\bf{t}}_{AI}}\}  + diag\{ {\bf{t}}_{MI}^*\} diag\{ {{\bf{t}}_{MI}}\}  + \sigma _R^2{{\bf{I}}_M}$.

Thereupon, the optimization problem of $\boldsymbol \psi$ is formulated as
\begin{align}
    \mathrm {(P6):}~\mathop {\max }\limits_{{\boldsymbol \psi}} \
          f_2({\boldsymbol \psi})
    \quad~~~ \text{s.t.} \
        {\boldsymbol \psi}^H \overline{\bf D} {\boldsymbol \psi}\leq P_{R,max},
\end{align}

Obviously, ${\boldsymbol \psi}$ not only appears in the numerator and denominator of the objective function, but also exists in the inverse matrix, which makes it difficult to solve this problem directly. Therefore, to deal with this problem, we first introduce a slack variable $\gamma $, which satisfies the condition
\begin{align}\label{gamma}
    \gamma &\geq({\boldsymbol \psi}^H
        \overline{\bf P}_{AIM}^H + \bar{\bf t}_{AM}^H) ({\boldsymbol \psi}^H \overline{\bf P}_{IM}{\boldsymbol \psi} {\bf h}_{IM,r}{\bf h}_{IM,r}^H \nonumber \\
        &+ {\bf R}_{NJ})^{-1}(\overline{\bf P}_{AIM}{\boldsymbol \psi} + \bar{\bf t}_{AM}).
\end{align}
And according to the nature of Schur complement, given matrix
\begin{align}
    &\qquad \qquad \qquad~~~~~~~~~ {\bf E}({\boldsymbol \psi}, \gamma) \\
    &=\left[
        {\begin{array}{cc}
        {{\boldsymbol \psi}^H \overline{\bf P}_{IM}{\boldsymbol \psi} {\bf h}_{IM,r}{\bf h}_{IM,r}^H
        + {\bf R}_{NJ}} &
            {\overline{\bf P}_{AIM}{\boldsymbol \psi} + \bar{\bf t}_{AM}}\\
        {{\boldsymbol \psi}^H
        \overline{\bf P}_{AIM}^H + \bar{\bf t}_{AM}^H} &
            {\gamma}
        \end{array}} \right], \nonumber
\end{align}
since ${{\boldsymbol{\psi }}^H}{\overline {\bf{P}} _{IM}}{\boldsymbol{\psi }}{{\bf{h}}_{IM,r}}{\bf{h}}_{IM,r}^H + {{\bf{R}}_{NJ}}$ is positive definite, inequality (\ref{gamma}) is equivalent to ${\bf{E}}({\boldsymbol{\psi }},\gamma )\succeq{\bf{0}}$.

Furthermore, after introducing the slack variable $\gamma$, a lower bound of $ln( {f_2}({\bf{\psi }}))$ can be obtained, i.e
\begin{align}
    g_2({\boldsymbol \psi}, \gamma) &= ln(1 + \frac{|\bar{\bf t}_{AIB}^H{\boldsymbol \psi} + \bar{l}_{AB}|^2}
         {|\overline{\bf P}_{MIB}^* {\boldsymbol \psi} + \bar{\bf t}_{MB}^*|^2 + |\overline{\bf P}_{IB}^* {\boldsymbol \psi}|^2 +  {\sigma}_B^2}) \nonumber \\
    &-ln(1+ \gamma) \nonumber \\
    &\leq ln(f_2({\boldsymbol \psi}))
\end{align}

Hence, the optimization problem can be transformed to
\begin{align}
    \mathrm {(P6-1):}~&\mathop {\max }\limits_{{\boldsymbol \psi, \gamma}} \quad \quad~~
          g_2({\boldsymbol \psi}, \gamma) \label{psi}\\
    &\ \text{s.t.} \quad~~~
        {\boldsymbol \psi}^H \overline{\bf D} {\boldsymbol \psi}\leq P_{R,max}, \tag{\ref{psi}{a}} \label{psia}\\
    & \quad ~~~~~~~ {\bf E}({\boldsymbol \psi}, \gamma) \succeq {\bf 0}.
        \tag{\ref{psi}{b}} \label{psib}
\end{align}
This problem is still non-convex and further treatment is needed. By referring to the lemma in \cite{alpha}, we find that for fixed point $\tilde{\alpha}$, $\tilde{\beta}$, the following inequalities hold,
\begin{align}\label{ab}
    ln(1 + \frac{|\alpha|^2}{\beta})
    &\geq ln(1+ \frac{|\tilde{\alpha}|^2}{\tilde{\beta}})
        + 2\frac{\Re\{ \tilde{\alpha}^*\alpha\}}{\tilde{\beta}} - \frac{|\tilde{\alpha}|^2}{\tilde{\beta}}\nonumber\\
    &-\frac{|\tilde{\alpha}|^2}{\tilde{\beta}(\tilde{\beta} + |\tilde{\alpha}|^2)}
        (\beta + |\alpha|^2),\\
    ln(1+\alpha) &\leq ln(1+ \tilde{\alpha}) + \frac{\alpha-\tilde{\alpha}}{1+ \tilde{\alpha}}.
\end{align}
Therefore, giving $\tilde{x} = \bar{\bf t}_{AIB}^H{\tilde{\boldsymbol \psi}} + \bar{l}_{AB}$ and $\tilde{y} = |\overline{\bf P}_{MIB}^* {\tilde{\boldsymbol \psi}} + \bar{\bf t}_{MB}^*|^2 + |\overline{\bf P}_{IB}^* {\tilde{\boldsymbol \psi}}|^2 +  {\sigma}_B^2$, it can be derived that
\begin{align}
    g_2({\boldsymbol \psi}, \gamma)
    &\geq
        ln(1+ \frac{|\tilde{x}|^2}{\tilde{y}}) + 2\frac{\Re\{\tilde{x}^*(\bar{\bf t}_{AIB}^H{\boldsymbol \psi}+ \bar{l}_{AB})\}}{\tilde{y}}-\frac{|\tilde{x}|^2}{\tilde{y}} \nonumber\\
    &-\frac{|\tilde{x}|^2}{\tilde{y}(\tilde{y} +|\tilde{x}|^2)}
        (|\overline{\bf P}_{MIB}^* {\boldsymbol \psi} + \bar{\bf t}_{MB}^*|^2 + |\overline{\bf P}_{IB}^* {\boldsymbol \psi}|^2
        \nonumber\\
    &+  {\sigma}_B^2 + |\bar{\bf t}_{AIB}^H{\boldsymbol \psi}
        + \bar{l}_{AB}|^2)
    - ln(1+ \tilde{\gamma}) - \frac{\gamma-\tilde{\gamma}}{1+ \tilde{\gamma}}
\end{align}

Moreover, for the un-convex constraint (\ref{psib}), we relax it by scaling the quadratic term in the matrix with respect to $\boldsymbol \psi$, as shown below.
\begin{align}
    &\qquad {\bf E}({\boldsymbol \psi}, \gamma) \succeq
        \overline{\bf E}({\boldsymbol \psi}, \gamma) \\
    &=\left[
        {\begin{array}{cc}
        {l {\bf h}_{IM,r}{\bf h}_{IM,r}^H
        + {\bf R}_{NJ}} &
            {\overline{\bf P}_{AIM}{\boldsymbol \psi} + \bar{\bf t}_{AM}}\\
        {{\boldsymbol \psi}^H
        \overline{\bf P}_{AIM}^H + \bar{\bf t}_{AM}^H} &
            {\gamma}
        \end{array}} \right], \nonumber
\end{align}
where $l = (2\Re\{{\boldsymbol \psi}^H \overline{\bf P}_{IM}\tilde{\boldsymbol \psi}\} - \tilde{\boldsymbol \psi}^H \overline{\bf P}_{IM}\tilde{\boldsymbol \psi} )$.

So far, the optimization problem of $\boldsymbol \psi$ is converted to
\begin{align}
    &\mathrm {(P6-2):}\nonumber\\
    &\mathop {\min }\limits_{{\boldsymbol \psi, \gamma}} \
        (|\overline{\bf P}_{MIB}^* {\boldsymbol \psi} + \bar{\bf t}_{MB}^*|^2 + |\overline{\bf P}_{IB}^* {\boldsymbol \psi}|^2 + |\bar{\bf t}_{AIB}^H{\boldsymbol \psi}
        + \bar{l}_{AB}|^2) \nonumber \\
    &\qquad \cdot\frac{|\tilde{x}|^2}{\tilde{y}(\tilde{y} +|\tilde{x}|^2)}+ \frac{\gamma}{1+ \tilde{\gamma}}
         -2\frac{\Re\{\tilde{x}^*\bar{\bf t}_{AIB}^H{\boldsymbol \psi}\}}{\tilde{y}}\nonumber \\
    &\ \text{s.t.} \qquad \qquad~~~
        (\ref{psia}),\quad
        \overline{\bf E}({\boldsymbol \psi}, \gamma) \succeq {\bf 0},
\end{align}
which is a SDP problem and can be solved by convex optimizing toolbox directly. And the specific steps of optimizing $\boldsymbol \psi$ with other variables fixed are provided in Algorithm 3.

\begin{algorithm}
\caption{Optimization Algorithm to (P6)}
\textbf{Input:}~initialize ${\boldsymbol \psi}^{(0)}$,$\gamma^{(0)}$, compute $g_2^{(0)}$, set $t=0$, convergence accuracy $\epsilon$.\\
\textbf{repeat:}
\begin{algorithmic}[1]
\item[1:]
    $t = t+1$.
\item[2:]
    Compute ${\widetilde x}$ and ${\widetilde y}$.
\item[3:]
    Update $\boldsymbol \psi^{(t)}$ and $\gamma^{(t)}$ by solving problem (P6-2).
\item[4:]
    Compute $g_2^{(t)}$.
\end{algorithmic}
\textbf{until:}~$g_2^{(t)} - g_2^{(t - 1)} < \epsilon$.\\
\textbf{Output:}~${\boldsymbol \psi}^{(t)}$.
\end{algorithm}

\subsection{Proposed low-complexity JO-PSM method}

\setcounter{TempEqCnt}{\value{equation}}
\setcounter{equation}{85}
\begin{figure*}
\begin{align}\label{f3}
    f_3({\boldsymbol \phi},{\boldsymbol \psi})
    &=  \underbrace{- c[|{\bf P}_{MIB}^* {\boldsymbol \theta} + {\bf t}_{MB}^*|^2
    + |{\bf t}_{AIB}^H{\boldsymbol \theta} + l_{AB}|^2]
    - ({\boldsymbol \theta}^H{\bf P}_{AIM}^H+ {\bf t}_{AM}^H){\bf F}_t^{-1}
    ({\bf P}_{AIM}{\boldsymbol \theta}+ {\bf t}_{AM})}_{g_2(\boldsymbol \theta)} \nonumber\\
    &+ \frac{2\Re\{\tilde{a}^*{\bf t}_{AIB}^H {\boldsymbol \theta}\}}{\tilde{b}}
    - c{\boldsymbol \psi}^H {\bf P}_{IB}^T {\bf P}_{IB}^* {\boldsymbol \psi}
    - e{\boldsymbol \psi}^H {\bf P}_{IM}{\boldsymbol \psi}
    +ln({\boldsymbol \psi}^H {\bf P}_{IMh} {\boldsymbol \psi} + \sigma_R^2)
\end{align}
\hrulefill
\end{figure*}
\setcounter{equation}{\value{TempEqCnt}}

The SO-PSM method proposed in the previous subsection optimizes $\boldsymbol \phi$ and $\boldsymbol \psi$ separately, and the optimization of $\boldsymbol \phi$ is based on SDR algorithm, therefore, as the number of IRS elements increases, the computational complexity of SO-PSM will become relatively high. Thus, in this subsection, we propose a low-complexity method for PSM optimization, namely JO-PSM, which optimizes $\boldsymbol \phi$ and $\boldsymbol \psi$ jointly.

In the derivation of JO-PSM method, we first find a lower bound of the objective function of problem (P4), and convert it into a problem that maximizes the lower bound, so as to find a suboptimal solution. According to (\ref{ab}), by setting $a = {\bf t}_{AIB}^H{\boldsymbol \theta} + l_{AB}$, $ b = ({\boldsymbol \theta}^H {\bf P}_{MIB}^T + {\bf t}_{MB}^T)({\bf P}_{MIB}^* {\boldsymbol \theta} + {\bf t}_{MB}^*)+ {\boldsymbol \psi}^H {\bf P}_{IB}^T {\bf P}_{IB}^* {\boldsymbol \psi} +\sigma_B^2$, the following inequality holds
\begin{align}
    {\overline R}_Bln2 &= ln(1 + \frac{|a|^2}{b})\nonumber\\
    &\geq ln(1+ \frac{|\tilde{a}|^2}{\tilde{b}})
        + 2\frac{\Re\{ \tilde{a}^*a\}}{\tilde{b}} - \frac{|\tilde{a}^2|}{\tilde{b}}\nonumber\\
    &- c(b + |a|^2),
\end{align}
where $\tilde{a} = {\bf t}_{AIB}^H \tilde{\boldsymbol \theta} + l_{AB}$, $ \tilde{b} = (\tilde{\boldsymbol \theta}^H {\bf P}_{MIB}^T + {\bf t}_{MB}^T)({\bf P}_{MIB}^* \tilde{\boldsymbol \theta} + {\bf t}_{MB}^*)+ \tilde{\boldsymbol \psi}^H {\bf P}_{IB}^T {\bf P}_{IB}^* \tilde{\boldsymbol \psi} +\sigma_B^2$, and $c = \frac{|\tilde{a}|^2}{\tilde{b}(\tilde{b}+ |\tilde{a}|^2)}$.

And due to the fact that $|{\bf I}_M + {\bf XY}| = |{\bf I}_N + {\bf YX}|$ for ${\bf X} \in \mathbb{C}^{M \times N}$ and ${\bf Y} \in \mathbb{C}^{N \times M}$, it can be derived that
\begin{align}\label{RMn}
    &\quad~~~~~~~~~~ {\overline R}_M*ln2 \nonumber\\
    &= ln|{\bf I}_{N_M} + ({\bf R}_{MR} +
        {\bf R}_{NJ})^{-1}({\bf P}_{AIM}{\boldsymbol \theta}
        + {\bf t}_{AM})\nonumber\\
    &\quad\cdot({\boldsymbol \theta}^H{\bf P}_{AIM}^H
        + {\bf t}_{AM}^H)|\nonumber\\
    & = ln|{\bf F}|-ln|{\bf R}_{MR} + {\bf R}_{NJ}|,
\end{align}
where ${\bf F}={\bf R}_{MR} + {\bf R}_{NJ} + ({\bf P}_{AIM}{\boldsymbol \theta}+ {\bf t}_{AM})({\boldsymbol \theta}^H{\bf P}_{AIM}^H+ {\bf t}_{AM}^H)$. For the first item in (\ref{RMn}), in accordance with the lemma mentioned in \cite{MM}, for fixed point $\widetilde{\boldsymbol\Sigma}$,
\begin{align}
    ln|{\boldsymbol\Sigma}| \leq ln|\widetilde{\boldsymbol\Sigma}| + tr[\widetilde{\boldsymbol\Sigma}^{-1}({\boldsymbol\Sigma}
    -\widetilde{\boldsymbol\Sigma})],
\end{align}
so setting ${\bf F}_t = {\tilde{\boldsymbol \psi}^H {\bf P}_{IM}\tilde{\boldsymbol \psi} {\bf h}_{IM,r}{\bf h}_{IM,r}^H + {\bf R}_{NJ}} + ({\bf P}_{AIM}\tilde{\boldsymbol \theta}+ {\bf t}_{AM})(\tilde{\boldsymbol \theta}^H{\bf P}_{AIM}^H+ {\bf t}_{AM}^H)$, we can derive that
\begin{align}
    ln|{\bf F}| &\leq ln|{\bf F}_t| + tr[{\bf F}_t^{-1}({\bf F}- {\bf F}_t)] \nonumber \\
    &= ln|{\bf F}_t| - N_M + tr[{\bf F}_t^{-1}{\bf F}] \nonumber\\
    &= ln|{\bf F}_t| - N_M + e{\boldsymbol \psi}^H {\bf P}_{IM}{\boldsymbol \psi}+ ({\boldsymbol \theta}^H{\bf P}_{AIM}^H+ {\bf t}_{AM}^H)\nonumber\\
    &\quad \cdot {\bf F}_t^{-1} ({\bf P}_{AIM}{\boldsymbol \theta}+ {\bf t}_{AM}) + tr[{\bf F}_t^{-1}{\bf R}_{NJ}],
\end{align}
where $e = tr[{\bf F}_t^{-1}{\bf h}_{IM,r}{\bf h}_{IM,r}^H]$.

Furthermore, for the second item in (\ref{RMn}), since ${\bf R}_{MR}$ and ${\bf R}_{AJ}$ are rank-one matrices, they can be represented as ${\bf R}_{MR} = \sigma_1^2 {\bf v}_1 {\bf v}_1^H$, ${\bf R}_{AJ} = \sigma_2^2 {\bf v}_2 {\bf v}_2^H$, then the second item of (\ref{RMn}) can be equivalently transformed as
\begin{align}
    &\quad~ln|{\bf R}_{MR} + {\bf R}_{NJ}|\nonumber\\
    &= ln|\sigma_1^2 {\bf v}_1 {\bf v}_1^H + \sigma_2^2 {\bf v}_2 {\bf v}_2^H + \sigma_R^2{\bf I}_{N_M}|\nonumber\\
    &= ln(\sigma_1^2 + \sigma_R^2) + ln(\sigma_2^2 + \sigma_R^2) +ln \sigma_R^{2(N_M-2)},
\end{align}
and referring to (\ref{rmr}), it is obvious that $\sigma_1^2 = {\boldsymbol \psi}^H{\bf P}_{IMh} {\boldsymbol \psi}$, with ${\bf P}_{IMh} = {\bf h}_{IM,r}^H{\bf h}_{IM,r}{\bf P}_{IM}$.

Therefore, with the above transformation, we can obtain a new objective function as (\ref{f3}), which is shown at the top of this page.

\setcounter{equation}{86}

Note that the new objective function is concave with respect to $\boldsymbol \phi$, however, due to the unit-modulus constraint, it is still hard to obtain the solution of $\boldsymbol \phi$. In this case, by referring to \cite{MM}, we
relax $g_3({\boldsymbol \theta})$, which contains all the quadratic terms of ${\boldsymbol \theta}$, as follows,
\begin{align}
    g_3(\boldsymbol \theta) &= -{\boldsymbol \theta}^H {\bf Q}_1 {\boldsymbol \theta} -2\Re\{ {\bf q}_1^H {\boldsymbol \theta}\} - \iota\\
    &\geq -{\boldsymbol \theta}^H(\lambda_{max}({\bf Q}_1) {\bf I}_M){\boldsymbol \theta}- \tilde{\boldsymbol \theta}^H(\lambda_{max}({\bf Q}_1) {\bf I}_M-{\bf Q}_1)\tilde{\boldsymbol \theta}
    \nonumber\\
    &+2\Re\{ \tilde{\boldsymbol \theta}^H(\lambda_{max}({\bf Q}_1) {\bf I}_M-{\bf Q}_1){\boldsymbol \theta}\}-2\Re\{ {\bf q}_1^H {\boldsymbol \theta}\} - \iota,\nonumber
\end{align}
where ${\bf Q}_1 = c{\bf P}_{MIB}^T{\bf P}_{MIB}^* + c{\bf t}_{AIB}{\bf t}_{AIB}^H + {\bf P}_{AIM}^H{\bf F}_t^{-1}{\bf P}_{AIM}$, ${\bf q}_1 = (c{\bf t}_{MB}^T{\bf P}_{MIB}^* + cl_{AB}^*{\bf t}_{AIB}^H + {\bf t}_{AM}^H{\bf F}_t^{-1}{\bf P}_{AIM})^H$ and $\iota = c{\bf t}_{MB}^T{\bf t}_{MB}^* + cl_{AB}^*l_{AB} + {\bf t}_{AM}^H{\bf F}_t^{-1}{\bf t}_{AM}$.

Specially, as ${\boldsymbol \phi}^H{\boldsymbol \psi} = 0$ and ${\boldsymbol \phi}^H{\boldsymbol \phi} = M-K$, then ${\boldsymbol \theta}^H(\lambda_{max}({\bf Q}_1) {\bf I}_M){\boldsymbol \theta} = \lambda_{max}({\bf Q}_1){\boldsymbol \psi}^H{\boldsymbol \psi} + \lambda_{max}({\bf Q}_1)(M-K)$. Thus, removing the constant terms and setting ${\bf Q}_2 = \lambda_{max}({\bf Q}_1){\bf I}_M + c{\bf P}_{IB}{\bf P}_{IB}^* + e{\bf P}_{IM}$, ${\bf q}_2 = (\tilde{\boldsymbol \theta}^H(\lambda_{max}({\bf Q}_1) {\bf I}_M-{\bf Q}_1) -{\bf q}_1^H + \tilde{a}^*/\tilde{b} {\bf t}_{AIB}^H )^H$ for brevity, the objective function can be transformed into
\begin{align}
    f_4({\boldsymbol \phi}, {\boldsymbol \psi})
    = 2\Re\{{\bf q}_2^H {\boldsymbol \theta} \}
    -{\boldsymbol \psi}^H {\bf Q}_2 {\boldsymbol \psi}
    + ln({\boldsymbol \psi}^H {\bf P}_{IMh} {\boldsymbol \psi} + \sigma_R^2).
\end{align}

Accordingly, the optimization problem turns into
\begin{align}
    \mathrm {(P7-1):}~&\mathop {\max }\limits_{{\boldsymbol \phi}, {\bold  \psi}} \qquad \quad
          f_4({\boldsymbol \phi},{\boldsymbol \psi})\nonumber\\
    &\ \ \text{s.t.}
        \quad {\boldsymbol \phi} = \mathcal{I}_M^{\bar{\mathbb{Q}}} {\boldsymbol \phi}, \ \
        {\boldsymbol \psi} = \mathcal{I}_M^{\mathbb{Q}} {\boldsymbol \psi}, \nonumber\\
    &\qquad \qquad
        |\phi_i| =1, \ i \in \mathbb{Q},\nonumber\\
    &\qquad \qquad
        {\boldsymbol \psi}^H{\bf D}{\boldsymbol \psi} \leq
        P_{R,max}.
\end{align}

At this point, $\boldsymbol \phi$ and $\boldsymbol \psi$ are no longer coupled, and the objective function of (P7-1) is a primary function with respect to $\boldsymbol \phi$, so the solution of $\boldsymbol \phi$ can be directly obtained as
\begin{align}\label{phiopt}
    \phi_i^{opt} =
     \begin{cases}
        0, \quad &i \in \mathbb{Q}, \\
        e^{j\ arg(q_{i})}, \quad \quad &i \notin \mathbb{Q},
     \end{cases}
\end{align}
where $q_i$ is the i-th element of ${\bf q}_2$.

As for $\boldsymbol \psi$, since ${\bf Q}_2$, ${\bf P}_{IMh}$ and ${\bf D}$ are diagonal matrices, setting $Q_{ii}$, $P_{ii}$ and $D_{ii}$ to denote the $i$-th element on the diagonal of these matrix respectively, the objective function with respect to $\boldsymbol \psi$ can be rewritten as
\begin{align}
    f_5({\boldsymbol \psi})
        &= \sum_{i\in \mathbb{Q}}(2\Re\{ q_i^* \psi_i\} - \psi_i^* Q_{ii} \psi_i)\nonumber\\
        &+ln(\sum_{i\in \mathbb{Q}} \psi_i^*P_{ii}\psi_i+\sigma_R^2).
\end{align}
which is un-convex due to the logarithmic term. Thus, we then use Jensen's inequality and (\ref{ab}) to scale the logarithmic term as follows
\begin{align}
     &\quad~ ln(\sum_{i\in \mathbb{Q}} \psi_i^*P_{ii}\psi_i+\sigma_R^2)
     \geq \sum_{i \in \mathbb{Q}}ln({\psi_i^*P_{ii}\psi_i+\sigma_R^2}/K)\nonumber\\
     &= \sum_{i \in \mathbb{Q}} (ln(1 + \psi_i^* P_{ir} P_{ir} \psi) - ln(\sigma_R^2/K))\nonumber\\
     &\geq \sum_{i\in \mathbb{Q}} (ln(1+|P_{ir}\tilde{\psi}|^2) + 2\Re\{\tilde{\psi}_i^* P_{ir} P_{ir} \psi_i\} \nonumber\\
     &\quad -\frac{|P_{ir} \tilde{\psi}|^2}{1+ |P_{ir} \tilde{\psi}|^2}\psi_i^* P_{ir}P_{ir} \psi_i - ln(\sigma_R^2/K)),
\end{align}
where $P_{ir} = (KP_{ii}/\sigma_R^2)^\frac{1}{2}$. For brevity, we set ${\bf q}_x = \{q_i + P_{ir} P_{ir} \tilde{\psi}_i \}$, ${\bf Q}_x = diag\{Q_{ii} + \frac{|P_{ir} \tilde{\psi}_i|^2}{1+ |P_{ir} \tilde{\psi}_i|^2} P_{ir} P_{ir} \}$,
${\bf D}_x = diag\{D_{ii}\}$, and ${\boldsymbol \psi}_x = \{\psi_i \}$, $i \in \mathbb{Q}$, and the optimization problem turns into
\begin{align}
    \mathrm {(P7-2):}~&\mathop {\min }\limits_{{\boldsymbol \psi}_x} \quad
         {\boldsymbol \psi}_x^H {\bf Q}_x {\boldsymbol \psi}_x
         -2 \Re\{{\bf q}_x^H {\boldsymbol \psi}_x\}
        \nonumber\\
    &\ \text{s.t.}\qquad \ \
         {\boldsymbol \psi}_x^H {\bf D}_x {\boldsymbol \psi}_x\leq P_{R,max}
\end{align}
This is a typical quadratically constrained quadratic programming (QCQP) problem, which can be solved by Lagrange multiplier method \cite{lag1,lag2}. Given $\mu$ as the Lagrange multiplier, the Lagrangian is
\begin{align}
    & L({\boldsymbol \psi}_x, \mu) \\
    &\quad = {\boldsymbol \psi}_x^H {\bf Q}_x {\boldsymbol \psi}_x
    -2 \Re\{{\bf q}_x^H {\boldsymbol \psi}_x\} + \mu ({\boldsymbol \psi}_x^H {\bf D}_x {\boldsymbol \psi}_x - P_{R,max}).\nonumber
\end{align}
It is obvious that when ${\boldsymbol \psi}_x = ({\bf Q}_x + \mu {\bf D}_x)^{-1} {\bf q}_x$, the lower bound of the Lagrangian, i.e. the objective function of the dual problem can be obtained, which is derived as
\begin{align}
    G(\mu) = -{\bf q}_x^H({\bf Q}_x + \mu{\bf D}_x)^{-1} {\bf q}_x -\mu P_{R,max}.
\end{align}
By exploring the complementary slackness condition of the constraint, it can be derived that if ${\bf q}_x^H {\bf Q}_x^{-2}{\bf D}_x {\bf q}_x \leq P_{R,max}$, $ \mu = 0$, otherwise, the optimal $\mu$ is the solution of the following problem
\begin{align}
    \mathrm {(P8):}~\mathop {\max }\limits_{\mu} \
        G(\mu) \quad
     \text{s.t.}\
         \mu > 0,
\end{align}
for which the optimal solution can be obtained by a one-dimensional search method.

Additionally, the specific steps of optimizing PSM with other variables fixed are provided in Algorithm 4.
\begin{algorithm}
\caption{Proposed JO-PSM method}
\textbf{Input:}~initialize ${\boldsymbol \phi}^{(0)}$, ${\boldsymbol \psi}^{(0)}$, compute ${\boldsymbol \theta}^{(0)}$, $({ \overline R _B} - {\overline R _M})^{(0)}$, set $t=0$, convergence accuracy $\epsilon$.\\
\textbf{repeat:}
\begin{algorithmic}[1]
\item[1:]
    $t = t+1$.
\item[2:]
    Compute ${\widetilde a}$, ${\widetilde b}$ and ${\bf F}_t$.
\item[3:]
     Compute ${\bf Q}_1$, $ {\bf q}_1$ and then compute ${\bf Q}_2$, $ {\bf q}_2$.
\item[4:]
    Update ${\boldsymbol \phi}^{(t)}$  by (\ref{phiopt}).
\item[5:]
    Compute ${\bf{q}}_x^H{\bf{Q}}_x^{-2}{{\bf{D}}_x}{{\bf{q}}_x}$, if ${\bf{q}}_x^H{\bf{Q}}_x^{ - 2}{{\bf{D}}_x}{{\bf{q}}_x} \le {P_{R,max}}$, then ${{\boldsymbol \psi}_x} = {\bf{Q}}_x^{ - 1}{{\bf{q}}_x}$. Otherwise, solve problem (P8) and then ${{\boldsymbol \psi }_x} = {({{\bf{Q}}_x} + \mu {{\bf{D}}_x})^{ - 1}}{{\bf{q}}_x}$. Update ${\boldsymbol \psi}^{(t)}$ based on ${\boldsymbol \psi}_x$.
\item[6:]
    Compute $({ \overline R _B} - {\overline R _M})^{(t)}$.
\end{algorithmic}
\textbf{until:}~$({ \overline R _B} - {\overline R _M})^{(t)} - ({ \overline R _B} - {\overline R _M})^{(t - 1)} < \epsilon$.\\
\textbf{Output:}~${\boldsymbol \theta} = {\boldsymbol \phi}^{(t)} +{\boldsymbol \psi}^{(t)}$.
\end{algorithm}


\section{Overall Algorithms and Complexity Analysis}
In the previous two sections, the optimization algorithms of beamforming and PSM were presented respectively. In this section, based on the algorithms of the above two sections, two joint design schemes of beamforming and PSM are given, and the computational complexity of the two schemes is analyzed.

\subsection{Proposed Max-SR-SOP}
By applying the receive beamformer and transmit beamformer in Section III, SO-PSM method in Section IV, Max-SR-SOP is proposed to maximize the SR by alternately optimizing the receive beamforming ${\bf v}_{BR}$, transmit beamforming $\bf v$, passive PSV $\boldsymbol \phi$ and active PSV $\boldsymbol \psi$.

The iterative idea of Max-SR-SOP algorithm is as follows: first, ${\bf v}_{BR}$, $\bf v$, $\boldsymbol \phi$ and $\boldsymbol \psi$ are initialized to feasible values; then, given fixed transmit beamforming $\bf v$ and PSV $\boldsymbol \phi$, $\boldsymbol \psi$, the receive beamforming vector ${\bf v}_{BR}$ is computed according to the generalized Rayleigh-Ritz theorem; given fixed receive beamforming ${\bf v}_{BR}$ and PSV $\boldsymbol \phi$, $\boldsymbol \psi$, ${\bf v}$ is updated by Algorithm 1; for given beamforming and $\boldsymbol \psi$, $\boldsymbol \phi$ is updated by Algorithm 2; then for given beamforming and $\boldsymbol \phi$, $\boldsymbol \psi$ is updated by Algorithm 3. The alternating iterations between the four variables will be repeated until the termination condition is satisfied, i.e. $({ \overline R _B} - {\overline R _M})^{(t)} - ({ \overline R _B} - {\overline R _M})^{(t - 1)} < \epsilon$, where $\epsilon $ denotes the iteration index. And then the optimized ${\bf v}_{BR}$, $\bf v$, $\boldsymbol \phi$ and $\boldsymbol \psi$ are finally obtained.

Specially, since the objective function value of problem (P1) is non-decreasing in each iteration, and the SR of the system is limited, Max-SR-SOP is guaranteed to converge to a feasible solution.

As for the computational complexity of Max-SR-SOP, it mainly originates from the iterations of the algorithm and the calculation of receive beamforming ${\bf v}_{BR}$, transmit beamforming $\bf v$, passive PSV $\boldsymbol \phi$ and active PSV $\boldsymbol \psi$. The computational complexity of receive beamforming design comes from matrix multiplications and eigenvalue decomposition, and its computational complexity can be expressed as
\begin{align}
{{\cal C}_{{{\bf{v}}_{BR}}}} &= {\cal O}(({N_B} + {N_M}){M^2}  + N_A^2{N_B} + N_B^3\nonumber\\
&+ (N_B^2 + N_M^2 + {N_B}{N_M} + {N_A}{N_B})M
).
\end{align}

The computational complexity of the transmit beamforming design, i.e., Algorithm 1, lies in solving the convex problem (P3-4) in each iteration. Therefore, by referring to \cite{complexity}, the computational complexity of updating $\bf v$ is given by
\begin{align}\label{compphi}
{{\cal C}_{\bf{v}}} = {\cal O}({L_{\bf{v}}}(N_B^3\ln (\frac{1}{{{\epsilon_0}}}))),
\end{align}
where ${L_{\bf{v}}}$ denotes the number of iterations of Algorithm 1 and $\epsilon_0$ is the calculation accuracy of CVX.

For the design of passive PSV, i.e. Algorithm 2, the computational complexity of solving problem (P5-2) is $\mathcal{O}({M^{6.5}}\ln (1/{\epsilon_0}))$, and that of Gaussian randomization is $\mathcal{O} (T{M^3})$ with $T$ representing the number of randomizations, thus, letting ${L_{\boldsymbol \phi} }$ denote the iterative number of Algorithm 2, the computational complexity of updating $\boldsymbol \phi$ is given by
\begin{align}
{{\cal C}_{\boldsymbol \phi} } = {\cal O}({L_{\boldsymbol \phi} }({M^{6.5}}\ln (\frac{1}{{{\epsilon_0}}}) + T{M^3})).
\end{align}

And for the design of active PSV, i.e. Algorithm 3, the computational complexity mainly lies in solving problem (P6-2) in each iteration, it can be derived that
\begin{align}\label{comppsi}
{{\cal C}_{\boldsymbol \psi} } =
{\cal O}({L_{\boldsymbol{\psi }}}(({M^3}N_M^{0.5} + {M^2}N_M^{2.5} + MN_M^{3.5})\ln (\frac{1}{{{\epsilon_0}}}))),
\end{align}
where ${L_{\boldsymbol{\psi }}}$ represents the iterative number of Algorithm 3.

Consequently, the computational complexity of Max-SR-SOP is
\begin{align}
{{\cal C}_{SOP}} = {L_{SOP}}({{\cal C}_{{{\bf{v}}_{BR}}}} + {{\cal C}_{\bf{v}}} + {{\cal C}_{\boldsymbol\phi} } + {{\cal C}_{\boldsymbol{\psi }}}),
\end{align}
where ${L_{SOP}}$ denotes the iterative number of Max-SR-SOP.

\subsection{Proposed Max-SR-JOP}
By using PSM optimization algorithm different from Max-SR-SOP, Max-SR-JOP is then proposed to reduce the computational complexity. In Max-SR-JOP, we apply the receive beamformer and transmit beamformer in Section III, JO-PSM method in Section IV, and the variables involved in the alternate iteration are reduced to three, which are receive beamforming ${\bf v}_{BR}$, transmit beamforming $\bf v$ and PSV $\boldsymbol \theta$, respectively.

The iterative idea of this algorithm is as fllows: first, ${\bf v}_{BR}$, $\bf v$, and $\boldsymbol \theta$ are initialized to feasible values; then, given fixed transmit beamforming $\bf v$ and PSV $\boldsymbol \theta$, the receive beamforming vector ${\bf v}_{BR}$ is computed according to the generalized Rayleigh-Ritz theorem; for given receive beamforming and $\boldsymbol \theta$, ${\bf v}$ is updated by Algorithm 1; next, for given beamforming, $\boldsymbol \theta$ is updated by Algorithm 4. The alternating iterations between the three variables will be repeated until the termination condition is satisfied, i.e. $({ \overline R _B} - {\overline R _M})^{(t)} - ({ \overline R _B} - {\overline R _M})^{(t - 1)} < \epsilon$, where $\epsilon $ denotes the iteration index. And then the optimized ${\bf v}_{BR}$, $\bf v$ and $\boldsymbol \theta$ are finally obtained. Similar as Max-SR-SOP, the convergence of Max-SR-JOP is also guaranteed as the objective function value is non-decreasing and the SR of system is limited.

The computational complexity is mainly derived from the iterations of the algorithm and the calculation of receive beamforming ${\bf v}_{BR}$, transmit beamforming $\bf v$, and PSV $\boldsymbol \theta$. The computational complexity of updating $\boldsymbol \theta$, i.e. Algorithm 4 mainly comes from matrix calculation, one-dimensional search and iteration. Letting the number of iterations be $L_{\boldsymbol \theta}$, the number of one-dimensional search be $L_{\mu}$, the computational complexity of Algorithm 4 is given by
\begin{align}\label{comptheta}
{{\cal C}_{\boldsymbol{\theta }}} = {L_{\boldsymbol{\theta }}}({M^3} + {N_B}{M^2} + {L_\mu }{K^3}).
\end{align}

As a result, the computational complexity of Max-SR-JOP can be expressed as
\begin{align}
{{\cal C}_{JOP}} = {L_{JOP}}({{\cal C}_{{{\bf{v}}_{BR}}}} + {{\cal C}_{\bf{v}}} + {{\cal C}_{\bf{\theta }}}),
\end{align}
with $L_{JOP}$ denoting the iterative number.

Observing (\ref{compphi}), (\ref{comppsi}) and (\ref{comptheta}), it is obvious that the highest computational complexity order of the Max-SR-SOP with respect to $M$ is 6.5, while that of the Max-SR-JOP is only 3. Hence, Max-SR-JOP has lower computational complexity, and as the number of IRS units increases, the gap between the two algorithms in computational complexity becomes more and more significant.

\section{Simulation Results}
In this section, the proposed Max-SR-SOP and Max-SR-JOP algorithms are numerically simulated to verify their feasibility and compare their performance. Unless otherwise stated, the simulation parameters are set as follows: $\beta  = 0.9$, $P_A = 30dBm$, ${P_{R,max}} = {P_M} = 20dBm$, ${N_A} = {N_B} = {N_M} = 5$, $M = 40$, $K = 2$ and $\sigma _B^2 = \sigma _M^2 = \sigma _R^2 =  - 40dBm$. Alice, HR-IRS, Bob and Mallory are located at (0m, 0m), (280m, 20m), (300m, 0m) and (150m, -20m) respectively. The path loss model is given as $g = {(c/4\pi fd)^2}$, where $c$ is the speed of light, $f$ denotes the carrier frequency of the signal, and $d$ represents the distance between the receiver and transmitter. For convenience, we directly set ${(c{\rm{/}}4\pi f)^2} = {10^{ - 2}}$ in the simulation. In addition, the convergence thresholds of all iterative algorithms are set to be $\epsilon=10^{-10}$.

\begin{figure}[t]
\setlength{\abovecaptionskip}{-5pt}
\setlength{\belowcaptionskip}{-10pt}
\centering
\includegraphics[scale=0.6]{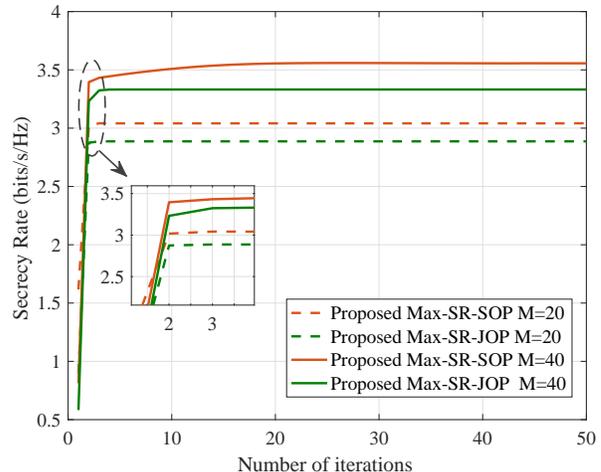}
\centering
\caption{Convergence of proposed algorithms}
\vspace{-0.2cm}
\end{figure}

First of all, we study the convergence behavior of the proposed Max-SR-SOP and Max-SR-JOP. The curves of SR versus number of iterations of these two algorithms are depicted in Fig.2. It is seen that both algorithms proposed in this paper can converge. In addition, as the number of HR-IRS phase shift elements increases, the SR of Max-SR-SOP and Max-SR-JOP will rise accordingly, while the number of iterations required to achieve convergence will also increase. By comparing the convergence curves of the two algorithms, it can be seen that when $M=20$, both algorithms can reach convergence after several iterations, and the convergence speed of Max-SR-JOP is slightly faster than that of Max-SR-SOP. When $M=40$, Max-SR-JOP can still converge within several iterations, but it takes tens of iterations for Max-SR-SOP to converge. This is because Max-SR-SOP has more parameters involved in the iteration and needs to go through more iterations to reach convergence. Besides, combined with the complexity expressions in the previous section, in each iteration, the highest order of the computational complexity of Max-SR-SOP with respect to $M$ is 6.5 and that of Max-SR-JOP is 3, from which it can be concluded that Max-SR-JOP has lower complexity than Max-SR-SOP. However, at the same time, the convergence value achieved by Max-SR-JOP is smaller than that achieved by Max-SR-SOP, the low complexity of Max-SR-JOP is achieved at the expense of some performance.

To measure the performance of the algorithms, in the following two pictures, we compare our proposed algorithms to the following benchmark schemes:

1) \textbf{No-IRS:} In this case, we set the PSM of HR-IRS to be zero, i.e. ${\boldsymbol \Theta}= {\bf 0}_{M \times M}$, and optimize beamforming vectors according to the algorithms in Section III.

2) \textbf{Random phase IRS:} In this case, we set the PSM of HR-IRS to be ${\boldsymbol \Theta}= diag\{ e^{j\mu_1}, ..., e^{j\mu_M}\}$, where $\mu_m, m=1,...,M$ is independently generated from $[0,2\pi)$, and then optimize beamforming vectors according to the algorithms in Section III.

3) \textbf{Passive IRS:} In this case, we set $K=0$ and $\mathcal{I}_M^{\mathbb{Q}}={\bf 0}_{M\times M}$, then optimize beamforming vectors and $\boldsymbol \phi$ according the algorithms in Section III and Algorithm 2.

4) \textbf{Passive IRS (${P_A+P_{R,max}}$):} In this case, we set the transmit power of Alice to be $P_A+P_{R,max}$, and then obtain the maximum SR by adopting the same settings as case 3).

\begin{figure}[t]
\setlength{\abovecaptionskip}{-5pt}
\setlength{\belowcaptionskip}{-10pt}
\centering
\includegraphics[scale=0.6]{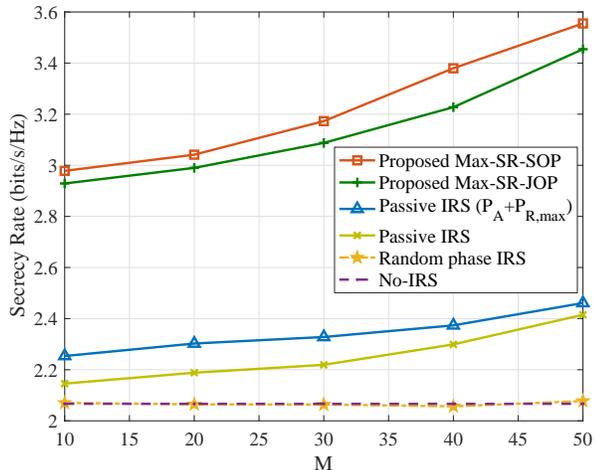}
\centering
\caption{SR versus the number of HR-IRS phase shift elements $M$}
\vspace{-0.2cm}
\end{figure}

Fig.3 depicts the SR versus the number of HR-IRS phase shift elements $M$ of the proposed algorithms and benchmark schemes. It can be seen from the figure that the scheme of HR-IRS-aided DM has a great performance improvement compared with passive IRS-aided DM and traditional DM, even under the same total system power budget. Moreover, the SR of Max-SR-JOP is slightly lower than that of Max-SR-SOP, which is consistent with the conclusion revealed in Fig.2, that is, although the proposed Max-SR-JOP has lower complexity, its performance is also slightly worse. Specially, when the phase shift of IRS is not optimized, the security performance of IRS-aided DM is roughly the same as that of traditional DM. This is reasonable because random-phase IRS may help to forward useful information as well as interference signals, which also reflects the importance of designing PSM. As for the trends of the curves, with the increase of the number of HR-IRS phase shift elements, the SR of the proposed two algorithms and passive IRS-aided DM also increases, and the SR of HR-IRS-aided DM increases slightly faster. Taking Max-SR-SOP as an example, when $M=20$, the SR of Max-SR-SOP is $38.8\%$ higher than that of passive IRS-aided DM, and when $M=50$, the SR is $47.2\% $ higher. 
\begin{figure}[t]
\setlength{\abovecaptionskip}{-5pt}
\setlength{\belowcaptionskip}{-10pt}
\centering
\includegraphics[scale=0.6]{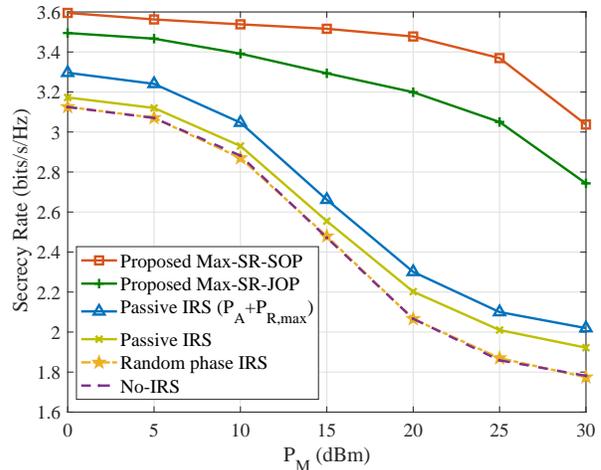}
\centering
\caption{SR versus the transmit power of Mallory $P_M$}
\vspace{-0.2cm}
\end{figure}

Fig.4 plots the SR versus the transmit power of Mallory $P_M$ for the proposed algorithms and benchmark schemes. Similar to Fig.3, permutations of all schemes in descending order of SR are Max-SR-SOP, Max-SR-JOP, Passive IRS ($P_A+ P_{R,max}$), Passive IRS, Random phase IRS, and the performance of Random phase IRS and No-IRS is roughly the same. It can be seen from this figure that as the power of malicious jamming increases, the SR performance of the six schemes decreases. When $P_M$ is small, the proposed Max-SR-SOP and Max-SR-JOP have similar anti-jamming performance, and when the $P_M$ increases, the performance gap between the two algorithms becomes obvious. In particular, according to the curves in Fig.4, the SR gap between the proposed Max-SR-SOP, Max-SR-JOP and other schemes increases first and then narrowed, which reflects that the SR of the proposed algorithms decreases with $P_M$ at a lower rate than that of passive IRS-aided DM and traditional DM. Therefore, it can be concluded that HR-IRS-aided DM has better anti-jamming performance.

\begin{figure}[t]
\setlength{\abovecaptionskip}{-5pt}
\setlength{\belowcaptionskip}{-10pt}
\centering
\includegraphics[scale=0.6]{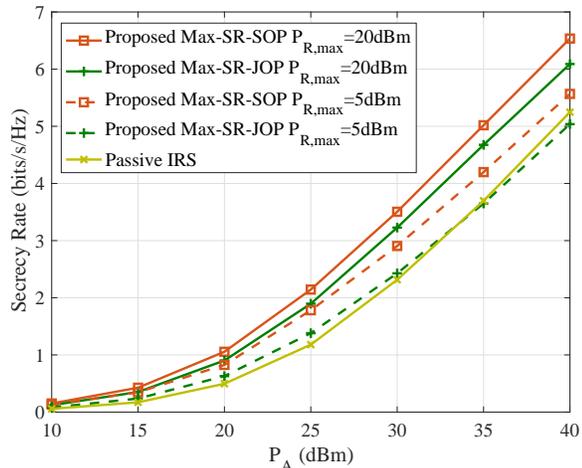}
\centering
\caption{SR versus the transmit power of Alice $P_A$}
\vspace{-0.2cm}
\end{figure}
To show the impact of the transmit power of Alice $P_A$ and the power budget of HR-IRS $P_{R,max}$, we plot the SR versus $P_A$ of the proposed algorithms under various power budget of HR-IRS, i.e. $P_{R,max}=5, 20 dBm$ in Fig.5. In addition, the curve of passive IRS-aided DM is also plotted as a benchmark. As seen from this figure, the SR of the proposed algorithms and passive IRS-aided DM increase with the increase of the transmit power of Alice, and with increase of the transmit power of HR-IRS, the SR of the proposed Max-SR-SOP and Max-SR-JOP will increase too.

When $P_{R,max} = 20dBm$, according to the simulation results, the performance of Max-SR-SOP and Max-SR-JOP is better than that of passive IRS-assisted DM scheme within the interval of ${P_A} \in [10,40](dBm)$. And when $P_{R,max}= 5dBm$, the SR curves of the proposed Max-SR-JOP and passive IRS-aided DM cross, that is, when ${P_A} \ge 35dBm$, the SR of Max-SR-JOP is lower than that of passive IRS-assisted DM, which is caused by the large gap between the transmit power and power budget of HR-IRS. Observing the expression of the transmit power of HR-IRS, i.e. (\ref{PR}) and the power constraint ${P_R} \le {P_{R,\max }}$, it can be seen that with the increase of $P_A$, in order to meet the power constraint, the amplification factor of the signal reflected by  active elements on HR-IRS will decrease. Thus, when $P_A$ is large and the power of HR-IRS $P_{R,max}$ is limited, the amplitude of the signal reflected by active elements on HR-IRS may be smaller than the amplitude of the incident signal, in this case, the security performance of the HR-IRS-aided DM will be worse than that of the passive IRS-aided DM. Therefore, when using HR-IRS, it should be provided with sufficient power based on the transmit power of Alice to ensure that HR-IRS can improve the security performance of the system.

\begin{figure}[t]
\setlength{\abovecaptionskip}{-5pt}
\setlength{\belowcaptionskip}{-10pt}
\centering
\includegraphics[scale=0.6]{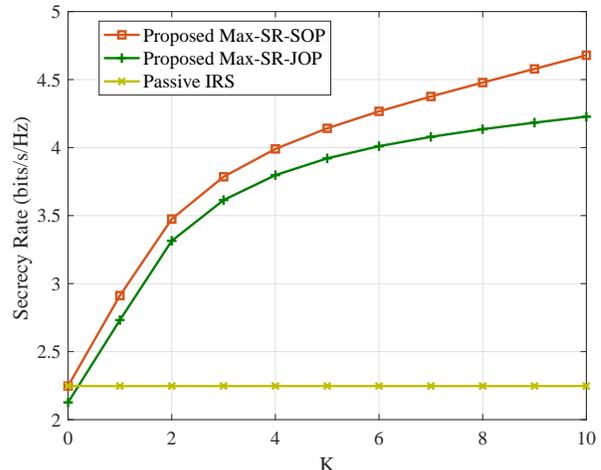}
\centering
\caption{SR versus the number of active units on HR-IRS $K$}
\vspace{-0.2cm}
\end{figure}

Fig.6 shows the SR versus the number of active units on HR-IRS $K$ of the proposed algorithms, and the curve of passive IRS-aided DM is also plotted as a benchmark. Since the proposed Max-SR-SOP adopts SDR algorithm to calculate the phase shift matrix corresponding to the passive units, which is identical to the algorithm used by the passive IRS-aided DM, when $K=0$, the SR of Max-SR-SOP and passive IRS-aided DM is the same. Meanwhile, the proposed Max-SR-JOP can obtain closed solution of PSM in each iteration at $K=0$, therefore has much lower complexity than the algorithm adopted by the passive IRS-aided DM, and correspondingly, the security performance is slightly lower. It is observed that with the increase of $K$, the SR of the two proposed algorithms increases. In particular, when the number of active elements on HR-IRS is small, increasing $K$ can improve the SR of the communication system in a larger proportion. Taking Max-SR-SOP algorithm as an example, the SR at $K=1$ is $29.6\%$ higher than that at $K=0$, while the SR at $K=10$ is only $2.2\%$ higher than that at $K=9$. This suggests that HR-IRS can greatly improve the security performance of the communication system by adding a small number of active units to passive IRS, thus balancing the performance and cost.

\section{Conclusion}
In this paper, we introduced HR-IRS into DM and investigated the design of beamforming and PSM in the HR-IRS-aided DM network with a malicious attacker. Under this system, an optimization problem aiming to maximize SR was established first, and then two iterative algorithms, namely Max-SR-SOP and Max-SR-JOP, were proposed to solve this problem. The main distinguish of the two algorithms is that Max-SR-SOP optimizes the active part and passive part of PSM separately, while the proposed Max-SR-JOP directly optimizes the whole PSM. The simulation results showed that the introduction of HR-IRS into the DM could effectively improve the security performance of the system. Moreover, the HR-IRS-aided DM scheme has better security and anti-jamming performance than passive IRS-aided DM scheme, when the  magnitude difference between the transmit power and the power budget of HR-IRS is not large. Meanwhile, Max-SR-SOP has better performance than Max-SR-JOP, while Max-SR-JOP has faster convergence speed and lower computational complexity.

\bibliographystyle{unsrt}
\bibliography{literature}

\end{document}